# Predicting Module-Lattice Reduction

Léo Ducas[1,2], Lynn Engelberts[1,3], and Paola de Perthuis[1]

[1] Centrum Wiskunde & Informatica, the Netherlands
{l.ducas,lynn.engelberts,paola.de.perthuis}@cwi.nl
[2] Leiden University, the Netherlands
[3] QuSoft, the Netherlands

**Abstract.** Is module-lattice reduction better than unstructured lattice reduction? This question was highlighted as 'Q8' in the Kyber NIST standardization submission (Avanzi et al., 2021), as potentially affecting the concrete security of Kyber and other module-lattice-based schemes. Foundational works on module-lattice reduction (Lee, Pellet-Mary, Stehlé, and Wallet, ASIACRYPT 2019; Mukherjee and Stephens-Davidowitz, CRYPTO 2020) confirmed the existence of such module variants of LLL and block-reduction algorithms, but focus only on provable worst-case asymptotic behavior.

In this work, we present a concrete average-case analysis of module-lattice reduction. Specifically, we address the question of the expected slope after running module-BKZ, and pinpoint the discriminant $\Delta_K$ of the number field at hand as the main quantity driving this slope. We convert this back into a gain or loss on the blocksize $\beta$: module-BKZ in a number field $K$ of degree $d$ requires an SVP oracle of dimension $\beta + \ln(|\Delta_K|/d^d)\beta/(d \ln \beta) + o(\beta/\ln \beta)$ to reach the same slope as unstructured BKZ with blocksize $\beta$. This asymptotic summary hides further terms that we predict concretely using experimentally verified heuristics. Incidentally, we provide the first open-source implementation of module-BKZ for some cyclotomic fields.

For power-of-two cyclotomic fields, we have $|\Delta_K| = d^d$, and conclude that module-BKZ requires a blocksize larger than its unstructured counterpart by $d-1+o(1)$. On the contrary, for all other cyclotomic fields we have $|\Delta_K| < d^d$, so module-BKZ provides a sublinear $\Theta(\beta/\ln \beta)$ gain on the required blocksize, yielding a subexponential speedup of $\exp(\Theta(\beta/\ln \beta))$.

## 1 Introduction

Module lattices were introduced in cryptography in 1996 with the NTRU cryptosystem [HPS98], and have since seen increasing interest driven by foundational results on their average-case hardness [SSTX09, LPR10, SS11, LS15]. In particular, they now underlie the security of three NIST post-quantum standards, ML-KEM, ML-DSA, and FN-DSA [NIS22], as well as a plethora of variants.

The central cryptanalytic tool to attack those cryptosystems is block lattice reduction, a term covering a variety of algorithms (including BKZ [Sch87, GN08b], slide [GN08a], and DBKZ [MW16]) that generalize LLL [LLL82] and offer a time-quality trade-off: roughly, block lattice reduction finds the shortest vector in a lattice in dimension $n$ up to an approximation factor of $\Theta(\sqrt{\beta})^{n/\beta}$ in time $2^{\Theta(\beta)}$. It operates by finding the shortest vector in *blocks* of dimension $\beta$, corresponding to projected sublattices of the $n$-dimensional lattice.

Until recently, lattice-reduction attacks were mostly oblivious to the module structure of the lattice, yet the potential of module variants of such algorithms was addressed in the documentation [ABD⁺21, Section 5.3, Q8] of Kyber (now standardized as ML-KEM) as part of its submission to the NIST standardization process. Specifically, question Q8 arises from the concern that the shortest-vector subroutine could benefit from a $d$ to $d^2$ speedup factor when applied to a module lattice over a cyclotomic field of degree $d$ [BNP17].

The discussion of Q8 [ABD⁺21, Sec 5.3, Q8] remarks that one could work over any subfield of the 512-th cyclotomic field used in Kyber, allowing to tune $d$ to any power of 2 between 1 and 256. The choice of a large $d$ is however highlighted as being at odds with various other speedups [AWHT16, LM18, Duc18]. More fundamentally, Q8 notes that such a module-BKZ algorithm, even when using the same blocksize as BKZ, may lead to a basis of slightly worse quality when applied to module lattices over cyclotomic fields of power-of-two conductors. Given that module-lattice analogs of block lattice reduction have now been developed [LPSW19, MS20], the following question can – and should – finally be addressed:

*Given the same shortest-vector oracle for lattices in dimension $\beta$, is module-BKZ better or worse than unstructured BKZ, and by how much?*

This question, relevant to all module-lattice-based schemes and not just Kyber, has not been fully answered in [LPSW19, MS20]: they study worst-case behavior of module-lattice analogs of LLL [LLL82] and slide reduction [GN08a], and are more concerned with the consequences of the existence of a fast short-vector oracle for module lattices of small rank over large number fields. Namely, these theoretical works present a worst-case to worst-case reduction from finding an approximately short vector in a module lattice to finding approximately short vectors in module lattices of smaller rank, up to some appropriate trade-off in the approximation factor that depends on the rank and quantities related to the number field. However, there is often a significant gap between the theoretical, worst-case understanding of block-reduction algorithms and their practical performance. This work therefore focuses on the average-case behavior of module-BKZ compared to BKZ, using a heuristic-based analysis. While [KK24] already experimented with module-LLL ($\beta = 2$) on the NTRU problem over power-of-two cyclotomic fields, they reported a negative result without providing a predictive analysis.

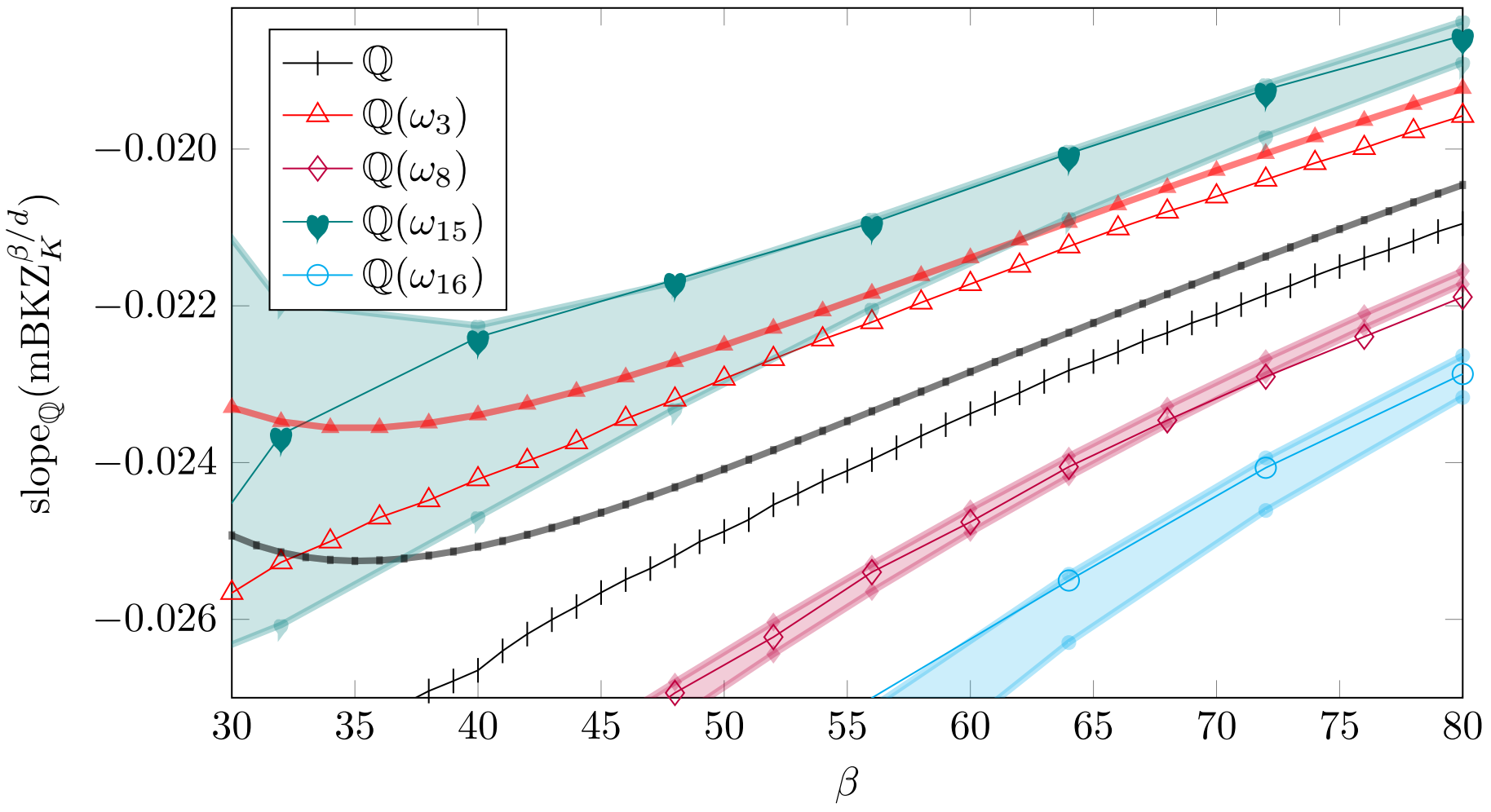

Fig. 1: Module-BKZ $\mathbb{Q}$-slope for several cyclotomic fields $K$. The case $K = \mathbb{Q}$ serves as a baseline for comparison with unstructured BKZ, under the general belief that BKZ is oblivious to the algebraic structure of module lattices. Experimentally measured slopes [.py] are represented by thin lines with large marks, and were averaged over 5 random lattices of dimension $rd = 240$. We progressively ran $5d$ tours for each multiple-of-$d$ blocksize $\beta$ to be close to convergence. Predictions [.py] consist of under- and overestimations, represented by thick lines with small marks and a filled region in between. For $\mathbb{Q}$ and $\mathbb{Q}(\omega_3)$, both predictions are too close to be distinguished.

### 1.1 Contributions

We propose a quantitative study of the practical performance of module-lattice analogs of block reduction. Specifically, we answer the aforementioned question using a heuristic analysis supported by extensive experiments, providing predictions on the quality of the output basis of module-BKZ as a function of the blocksize $\beta$, measured by the *slope* of the so-called basis profile. As visible in Figure 1, these predictions seem to fit reasonably well with our experiments.[4] A small gap remains, which may be in part due to head and tail phenomena [YD17, BSW18] unaccounted for by the slope model (the Geometric Series Assumption).

More precisely, we show that the 'equivalent' blocksize[5] $\beta_{\mathrm{eq}}$ of module-BKZ, required to achieve the same slope as BKZ with blocksize $\beta$, is asymptotically [.py] (see Heuristic Claim 3 for the more refined expression used to generate Figure 2):

$$\beta_{\mathrm{eq}} = \beta + \ln\left(\frac{|\Delta_K|}{d^d}\right)\frac{\beta}{d\ln\beta}(1 + o(1)) + d - 1 + o(1) \tag{1}$$

[4] We also conducted experiments confirming the belief that BKZ is oblivious to the structure of the embedded module lattices [.py].

[5] For easier comparison, blocksize is measured as the *dimension* (or $\mathbb{Q}$-rank) of the lattice, not the $K$-rank.

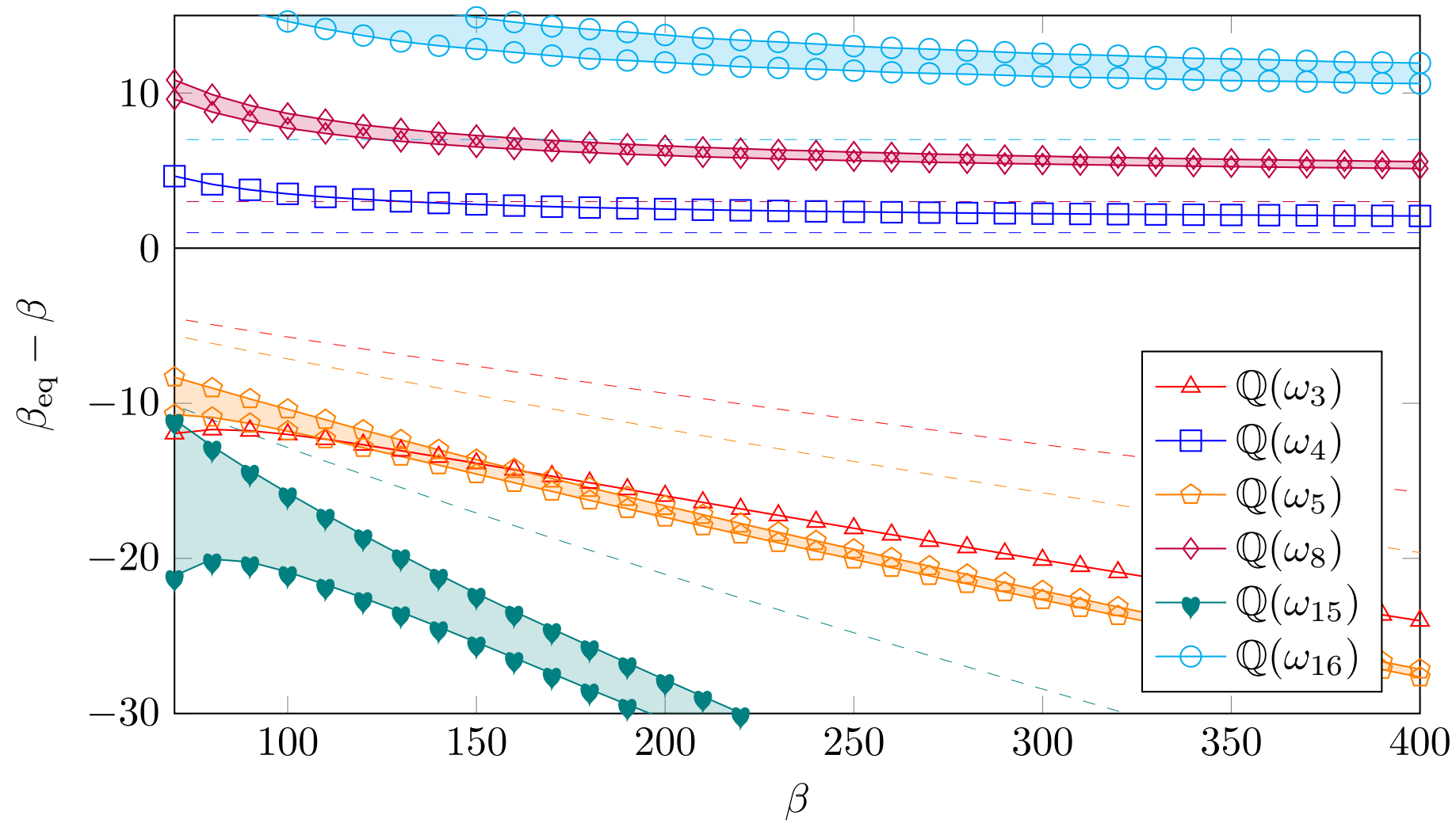

Fig. 2: Predictions for the difference $\beta_{\text{eq}} - \beta$ of blocksizes for $\text{mBKZ}_K^{\beta_{\text{eq}}/d}$ to reach the same slope as unstructured $\text{BKZ}^\beta$ for several cyclotomic fields $K$. Concrete predictions [.py] are represented by lines with large marks and a filled region in between. Asymptotic predictions from Heuristic Claim 3 [.py] are represented by dashed lines.

where $d$ denotes the degree of the underlying number field $K$ and $\Delta_K$ its discriminant. The $d - 1 + o(1)$ term of the asymptotic formula plays a role in the $|\Delta_K| = d^d$ case, which for cyclotomic fields corresponds to power-of-two conductors. These asymptotic results are merely intended as a summary of our analysis, as we also provide concrete predictions using explicit formulas and prediction scripts in `Python`. Equation (1) shows that, for a fixed number field, the gain or loss on the blocksize is barely sublinear with a constant depending on the discriminant $\Delta_K$, and our concrete estimates show it is quite substantial in practice, see Figure 2. This figure also shows that the asymptotic summary is not precise enough to be used as an approximation for concrete security estimates.

Incidentally, we provide an implementation of module-BKZ based on `fplll` [dt23] and `G6K` [ADH+19] for cyclotomic fields, at least up to conductor 16. While it is in principle more general, it fails in certain cases due to technical limitations that we discuss later on. While our implementation is far from optimized, it already allows experimentation with the behavior of module-BKZ beyond the strict focus of this paper, and should be useful for answering some of the open questions listed below. It is available at `https://github.com/lducas/mBKZ/`, and our paper systematically points to the relevant bits of our code using the symbol [.py].

### 1.2 Cryptanalytic Impact

Let us now outline the potential implications of this study for cryptographic schemes based on module lattices. First of all, our observations are not bound to the scheme's underlying field, and apply to an arbitrary subfield of it. In particular, for a scheme over a cyclotomic field $\mathbb{Q}(\omega_s)$ of conductor $s$, an attacker can work over any $\mathbb{Q}(\omega_c)$ for $c$ dividing $s$. For example, for Kyber, an attacker has the freedom to work over cyclotomic fields of degree $d = 2^k$ for $k \leqslant 8$. A larger $d$ should allow them to obtain speedups in the shortest-vector oracle, but also restricts them to use a blocksize multiple of this larger value of $d$.

In the case of power-of-two cyclotomic fields, which are relevant to the new NIST standards ML-KEM, ML-DSA, and FN-DSA [NIS22], the main asymptotic term $\ln(|\Delta_K|/d^d)\beta/(d \ln \beta)$ disappears as the discriminant of such a field $K$ satisfies $|\Delta_K| = d^d$. The remaining term is $d - 1 + o(1)$, and Figure 2 shows a slow[6] convergence from above. This explains the disappointing performance of Karenin and Kirshanova's module-LLL implementation for overstretched NTRU parameters [KK24]. More importantly, this confirms and quantifies the discussion of Kyber's Q8 [ABD+21, Sec 5.3, Q8], which already suggested module-BKZ might need an increased blocksize in order to match the output quality of unstructured BKZ. Nevertheless, quite a bit of work remains to settle the question of whether module-BKZ can slightly outperform BKZ in this context, see the following Open Questions 1, 2, and 3.

On the contrary, for other cyclotomic fields, especially those whose conductor has one or more small odd prime factors, we predict a rather substantial gain on the slope, as illustrated in Figure 2. In fact, we predict that each additional odd prime factor leads to a further slope gain: for instance, compare $\mathbb{Q}(\omega_{15})$ to $\mathbb{Q}(\omega_3)$ and $\mathbb{Q}(\omega_5)$ in Figure 2. The slope gain induced by odd prime factors is relevant as popular alternatives to power-of-two conductors are conductors of the form $2^i \cdot 3^j$ [BDF18, EFG+22, EWY23]. In that case, module-BKZ over the third cyclotomic field is predicted to gain 20 more dimensions on the blocksize at NIST security level 1 ($\beta \approx 380$).

There are more schemes using conductors of the form $2^i \cdot 3^j$ [PFH+17, LS19, HWK+25, Kpq25]. These include an intermediate parameter set of Falcon when it was submitted to the first round of the NIST PQC standardization process, and one of the two selected Korean post-quantum PKE/KEM standards NTRU+. The Homomorphic Encryption library `HElib` also supports general conductors, and a set of parameters with a multiple-of-5 conductor was highlighted in [HS14]. However, these schemes use the Euclidean norm defined by the coefficient embedding rather than the canonical embedding, and the impact of our observations would require special consideration (see Open Question 5).

Lastly, our analysis might narrow down avenues to construct more efficient schemes based on module-LIP [DPPvW22]. The LIP framework [DvW22] was designed to harness the decoding capabilities of dense lattices in cryptography, but competitive proposals likely require a module structure. One would naturally turn to the algebraic construction of remarkable lattices, such as the construction [BF00] of the Leech lattice as an ideal in a cyclotomic field of conductor $35, 39, 52, 56$, or 84. Worse, one could be tempted to use Martinet's construction [Mar78] of asymptotically dense lattices based on towers of number fields of bounded discriminants.

[6] For example, for $\mathbb{Q}(\omega_{16})$, we have $d - 1 = 7$, but $\beta_{eq} - \beta \in [11.2, 12.6]$ at $\beta = 400$.

### 1.3 Open Questions

Although the prediction and experimental analysis of the module-BKZ slope highlight its potential advantages, this study alone does not suffice to precisely quantify the cost of attacks based on module-BKZ. We list several future directions, including continued investigation of Q8 [ABD+21, Sec 5.3, Q8]. While our implementation may help answering some of those questions, a proper API for module-lattice reduction would be beneficial.

1. **HKZ Profile, Tails, and Dimensions for Free.** Just as the module structure affects the BKZ slope, we expect it to affect the profile of module-HKZ reduction as well. Predicting HKZ shapes in a similar manner would allow replacing the Geometric Series Assumption by its tail-adapted refinement [AD21]. Perhaps more critically, this change of shape should also affect the number of dimensions for free in the shortest-vector subroutine [Duc18], thereby slowing down or accelerating the shortest-vector oracle, in addition to altering the slope.

2. **Profile Simulation.** This study is limited to the slope of BKZ after convergence, i.e., after many tours. In practice, cryptanalysts are more aggressive and run only a single or a few tours, progressively increasing the blocksize [AWHT16]. The fine-tuning of attacks and security estimates then resorts to BKZ simulators [CN11,BSW18,XWW+24], which should be adapted to module-BKZ. In particular, one may question how fast module-BKZ converges compared to unstructured BKZ.

3. **Advanced Lattice Sieving with Cyclotomic Symmetries.** Although our results suggest that module-BKZ performs worse than BKZ for power-of-two cyclotomic fields, the part that we model as shortest-vector oracle calls relies in practice on subroutines that may benefit from a cyclotomic structure. Indeed, speedups and memory savings have been demonstrated for sieving over cyclotomic ideal lattices [BNP17]. However, these results were obtained for a rather naive sieving algorithm [MV10], and it is far from clear whether the same methods would combine well in practice with improved sieving techniques based on locality-sensitive hashing [Laa16, BDGL16] and other practical tricks used by the fastest known sieving implementations, such as progressive sieving and the dimensions-for-free technique [LM18, Duc18, ADH+19, DSvW21].

4. **Solving Cryptographic Module-Lattice Problems.** Another important open question is whether the observed slope gain translates into faster algorithms for solving module-SIS, module-LWE, module-LIP, and NTRU. A preliminary proof of concept (see Appendix C) answers this question positively, but precise predictions would require profile simulation (Open Question 2) and a probabilistic analysis of secret recovery [DDGR20, PV21].

5. **Coefficient Embedding.** Our predicted and observed slope gains are obtained using the canonical embedding to define the geometry of cyclotomic fields. While it is algebraically more natural to use the canonical embedding, some schemes [HS14, PFH+17, LS19, HWK+25] use the coefficient embedding instead. We note that the distortion to go from one embedding to the other depends only on the number field (in fact, no distortion occurs for power-of-two cyclotomic fields), and is constant with growing module rank $r$. On the contrary, our slope gain leads to a gain on the first basis vector's length that grows exponentially with $r$ for a fixed blocksize. Hence, if $r$ is large enough, it will eventually be beneficial to apply module-BKZ using the canonical embedding even when the targeted scheme uses a different embedding. Nevertheless, some study is required to determine the exact break-even point and make concrete predictions for this setup.

6. **Shortest-Vector versus Densest-Ideal Oracle.** From a theoretic perspective [MS20, KK24], it would be more natural for module-BKZ to use an oracle for finding a *densest ideal* rather than a shortest vector. This raises two questions: to what extent the slope would improve, and how such an oracle could be realized reasonably efficiently compared to the best shortest-vector oracles [BDGL16, Duc18]. The former question boils down to establishing a Gaussian Heuristic for the algebraic norm.

### 1.4 Technical Overview

We start by briefly recalling notions from lattice reduction: the quality of a lattice basis $(\mathbf{z}_1, \ldots, \mathbf{z}_n) \subseteq \mathbb{R}^{n\times n}$ is measured by its *profile*, namely the sequence $\ln \|\mathbf{z}_1^*\|, \ldots, \ln \|\mathbf{z}_n^*\|$ of logarithmic norms of its Gram-Schmidt vectors. The sum of those logarithms is an invariant of the lattice, the logarithm of its determinant. This profile is typically decreasing in $i$, and the flatter it is, the better. Since $\mathbf{z}_1^* = \mathbf{z}_1$, a reduced basis provides in particular a short nonzero lattice vector. Note that a vector $\mathbf{b} \in K^r$ can naturally be viewed as a vector $\mathbf{z} \in \mathbb{R}^{rd}$ through the canonical embedding, allowing us to refer to its length $\|\mathbf{b}\|$ as the Euclidean norm of $\mathbf{z}$. See Section 2 for more details.

*Two Enlightening Cases.* Consider an $\mathcal{O}_K$-module lattice over the fourth cyclotomic field $K = \mathbb{Q}(\imath)$, for $\mathcal{O}_K$ its ring of integers. Having found a somewhat short vector $\mathbf{b}_1$, we want to use it to perform module-structured lattice reduction. Naturally, one would set $\mathbf{b}_2 := \imath\mathbf{b}_1$.[7] It is always the case that $\mathbf{b}_1$ is orthogonal to $\imath\mathbf{b}_1$, hence $\|\mathbf{b}_1^*\| = \|\mathbf{b}_2^*\|$. The profile may look perfectly flat locally, but this constraint actually makes the global profile less flat: because $\|\mathbf{b}_2^*\|$ is a bit larger than it would have been for an unstructured reduction, more of the determinant has been consumed, lowering the rest of the profile. This discussion implicitely assumes that the length of $\mathbf{b}_1$ would be the same in the structured and unstructured cases, a matter we will discuss below.

[7] Other valid choices are $\mathbf{b}_2 := (k + \imath)\mathbf{b}_1$ for $k \in \mathbb{Z}$, but will not change $\mathbf{b}_2^*$.

If we instead consider the third cyclotomic, $\mathbb{Q}(\omega_3)$, the situation is rather different. In this case, $\mathbf{b}_1$ and $\mathbf{b}_2 := \omega_3 \mathbf{b}_1$ always form an angle of $\pi/3$, so $\|\mathbf{b}_1^*\| = \sqrt{4/3} \cdot \|\mathbf{b}_2^*\|$, and $\mathbf{b}_2^*$ is significantly shorter than what it would be in an unstructured reduction, making the profile locally more inclined, but globally flatter.

*The General Case.* Consider an $\mathcal{O}_K$-module lattice $\mathcal{M}$ for some number field $K$. Writing $\mathbf{b}_1$ for the first vector in a basis of $\mathcal{M}$, its length is predicted by the Gaussian Heuristic in the unstructured case, and backed by more formal theorems [Rog56, Söd11, Che13, LN20] using a careful definition of random lattices. Fortunately, an adaptation of those theorems has recently been proven for module lattices [GSVV24], which we use as a module-lattice Gaussian Heuristic in our analysis. It predicts that the shortest vector of a random cyclotomic module lattice is only barely larger than in a random unstructured lattice.

If $K$ is an imaginary quadratic number field, we further observe that the rank-1 module lattice $\mathbf{b}_1\mathcal{O}_K$ is always a scaled rotation of $\mathcal{O}_K$ itself, and the quality of the first part of this basis is therefore directly related to the density of $\mathcal{O}_K$ as a lattice, or, in algebraic terms, to the absolute discriminant $|\Delta_K|$.

The situation is a bit more complex beyond imaginary quadratics, where $\mathbf{b}_1\mathcal{O}_K$ need not be a scaled rotation of $\mathcal{O}_K$: it gets *skewed*. Yet, our heuristic analysis below (adapted from [DvW21, Lemma 4.4]) shows that as we model $\mathbf{b}_1$ as following a spherical distribution, the skewness quickly decreases as the rank increases, making this concern asymptotically irrelevant, and concretely controlled. We observe less average skewness experimentally than predicted by our model, and provide an explanation of why this model is not perfectly accurate. As we have no better model to offer, we translate this into an interval with one end corresponding to no skewness and the other corresponding to the spherical-model estimate. Luckily, the consequence of that uncertainty is quantitatively mild, fading away as the blocksize grows.

Another complication can happen, namely that $\mathbf{b}_1\mathcal{O}_K$ does not capture all the module-lattice points in $\mathbf{b}_1 K$. What we really want to construct as the first rank-1 module is $\mathbf{b}_1 K \cap \mathcal{M}$, where $\mathcal{M}$ is the module lattice at hand. Here again, a heuristic analysis involving the Dedekind zeta function (in a fashion similar to [ABD16, DPPvW22, DvW21]) allows modeling the distribution of the *index* of $\mathbf{b}_1\mathcal{O}_K$ in $\mathbf{b}_1 K \cap \mathcal{M}$. Again, this model appears to be an overestimate compared to the index encountered experimentally, for an explainable reason, without an obvious fix, but luckily again the uncertainty it leaves is inconsequential.

This gives us four terms driving the slope of module-BKZ: the module-lattice Gaussian Heuristic, the discriminant, the skewness, and the index. Putting it all together, we conclude with a concrete slope prediction, and an asymptotic analysis of the gain or loss in terms of the blocksize.

*Dense Sublattices.* From the discussion above, we see that what can make module-BKZ more performant than vanilla BKZ is its ability to find not just short vectors, but dense sublattices (which happen to be ideals). Interestingly, this advantage is not only about the ability to find them, but also about their mere existence. Indeed, a recent work [DL25] predicted the slope of a variant of BKZ based on a densest-sublattice oracle for unstructured lattices, and concluded that the densest sublattices of random unstructured lattices are not dense enough to significantly improve upon vanilla BKZ.

## 2 Preliminaries

**Notation.** The set of positive (rational) integers is denoted by $\mathbb{N}$. For integers $x, x' \in \mathbb{Z}$, we define $[\![x; x']\!] := \{x, x+1, \ldots, x'\}$. We denote Euler's totient function by $\phi$: for $x \in \mathbb{N}$, $\phi(x)$ equals the number of integers in $[\![1; x]\!]$ coprime to $x$.

We write $x \sim \mathcal{D}$ to denote that $x$ is sampled from the distribution $\mathcal{D}$. We will use $\mathbb{E}_{x\sim\mathcal{D}}[f(x)]$ to denote the expected value of $f(x)$ for $x \sim \mathcal{D}$. Whenever we do not specify the distribution, we implicitly refer to the (unknown) distributions encountered during the BKZ or module-BKZ algorithms (see Remark 1).

Throughout this paper, we use the symbol [.py] for external (clickable) links to the corresponding part of our code where available.

### 2.1 Lattice Background

A *(Euclidean) lattice* in $\mathbb{R}^n$ is a set of the form $\mathcal{L} = \mathcal{L}(\mathbf{B}) := \mathbf{B}\mathbb{Z}^k$ for some $\mathbf{B} \in \mathbb{R}^{n\times k}$ with linearly independent columns. We call $\mathbf{B}$ a basis for $\mathcal{L}$, and say that $\mathcal{L}$ has dimension $k$. Given a basis $\mathbf{B}$, we refer to its *GSO* as the set of vectors obtained through Gram-Schmidt orthogonalization.

An important invariant of a lattice $\mathcal{L}$ is its *first minimum*, denoted by $\lambda_1(\mathcal{L})$ and defined as $\lambda_1(L) := \inf\{\|\mathbf{x}\| : \mathbf{x} \in \mathcal{L}\backslash\{\mathbf{0}\}\}$, where $\|\cdot\|$ is the Euclidean norm. The task of finding a nonzero lattice vector of minimal length is known as the *Shortest Vector Problem* (SVP). In this work, we will assume that we have an SVP oracle at our disposal, i.e., an algorithm that gives us a shortest nonzero lattice vector when given a basis of a lattice.

Another important invariant of a lattice $\mathcal{L}$ is its *determinant* $\det_{\mathbb{Q}}(\mathcal{L})$, also often called its volume (or, more accurately, its covolume). It is defined as $\det_{\mathbb{Q}}(\mathcal{L}) := \sqrt{|\det(\mathbf{B}^T\mathbf{B})|}$ for a basis $\mathbf{B}$ of $\mathcal{L}$; yet its value is independent of the basis in consideration. When $\mathcal{L}$ has dimension $k$, we define the *normalized* lattice $\mathcal{L}^{(1)} = \mathcal{L}/\det_{\mathbb{Q}}(\mathcal{L})^{1/k}$, so that $\mathcal{L}^{(1)}$ has determinant equal to 1.

*Remark 1 (On random lattices).* Our work studies the performance of module-BKZ on *random lattices*, which involves analyzing various related random lattices encountered during the algorithm. It is therefore worth mentioning what we mean with 'random' here. It is well known that using the Haar measure, one can formally define a uniform distribution over the set of $n$-dimensional lattices of unit determinant [Sie45]. However, as is the case for unstructured BKZ, the distribution of the random lattices encountered in our analysis of module-BKZ is not well understood. The notion of random lattice in this work is therefore not always explicitly defined, and refers to its actual distribution induced by module-BKZ. Nevertheless, we circumvent this lack of understanding in a similar manner as is done in analysis of unstructured BKZ: we approximate these distributions using heuristics that we verify experimentally.

### 2.2 Algebraic Background

Let $K$ be a number field of degree $d = [K : \mathbb{Q}]$, i.e., $K \cong \mathbb{Q}[X]/P(X)$ for some irreducible polynomial $P \in \mathbb{Q}[X]$ of degree $d$. $K$ admits $d$ distinct *embeddings* into $\mathbb{C}$, i.e., injective field homomorphisms from $K$ to $\mathbb{C}$. Each of these embeddings corresponds to evaluating elements of $K$ at one of the roots of $P$ in $\mathbb{C}$. We have $d = d_\mathbb{R} + 2d_\mathbb{C}$, where $d_\mathbb{R}$ denotes the number of real embeddings (corresponding to the roots of $P$ in $\mathbb{R}$) and $d_\mathbb{C}$ the number of complex embeddings, up to conjugation. We denote the set of real embeddings by $\mathcal{E}_\mathbb{R}$, the $d_\mathbb{C}$ embeddings corresponding to roots with strictly positive imaginary part by $\mathcal{E}_\mathbb{C}^+$, and their conjugates by $\overline{\mathcal{E}_\mathbb{C}^+}$. Altogether, the set $\mathcal{E}$ of all $d$ embeddings of $K$ decomposes as $\mathcal{E} = \mathcal{E}_\mathbb{R} \uplus \mathcal{E}_\mathbb{C}^+ \uplus \overline{\mathcal{E}_\mathbb{C}^+}$.

We define the ring $K_\mathbb{R}$ as $K \otimes_\mathbb{Q} \mathbb{R}$. Note that we have a ring isomorphism $K_\mathbb{R} \cong \mathbb{R}^{d_\mathbb{R}} \times \mathbb{C}^{d_\mathbb{C}}$. We define the field *trace* by $\mathrm{Tr}\colon K_\mathbb{R} \to \mathbb{R}, x \mapsto \sum_{j=1}^{d} \sigma_j(x)$, inducing an inner product on $K_\mathbb{R}$ given by $\langle x, y\rangle := \mathrm{Tr}(x\overline{y})$ for $x, y \in K_\mathbb{R}$. This trace inner product is extended to $K_\mathbb{R}^m$ by defining $\langle \mathbf{x}, \mathbf{y}\rangle := \sum_{i=1}^{m} \langle x_i, y_i\rangle = \sum_{i=1}^{m} \mathrm{Tr}(x_i\overline{y}_i)$ for $\mathbf{x} = (x_1, \ldots, x_m), \mathbf{y} = (y_1, \ldots, y_m) \in K_\mathbb{R}^m$. In particular, this yields a geometric norm $\|\cdot\|\colon \mathbf{x} \mapsto \sqrt{\langle \mathbf{x}, \mathbf{x}\rangle}$. We also define $\langle \mathbf{x}, \mathbf{y}\rangle_K := \sum_{i=1}^{m} x_i\overline{y}_i \in K_\mathbb{R}$.

Besides the trace norm $\|x\| = \sqrt{\langle x, x\rangle}$ of $x \in K_\mathbb{R}$, we consider the *algebraic norm* $\mathrm{N}(x) := \prod_{j=1}^{d} \sigma_j(x)$. By the arithmetic-geometric inequality, we have $\sqrt{d}\mathrm{N}(x)^{1/d} \leqslant \|x\|$. For $\mathbf{x} \in K_\mathbb{R}^m$, we write $\mathrm{N}(\mathbf{x})$ as shorthand for $\mathrm{N}(\langle \mathbf{x}, \mathbf{x}\rangle_K)^{1/2}$.

We write $\mathcal{O}_K$ for the *ring of integers* of $K$, which consists of the elements $x \in K$ such that $Q(x) = 0$ for some monic polynomial $Q \in \mathbb{Z}[X]$. In particular, $\mathcal{O}_K$ is a free $\mathbb{Z}$-module of rank $d$, i.e., it is the set of all $\mathbb{Z}$-linear combinations of a basis $(x_1, \ldots, x_d)$ in $\mathcal{O}_K$ (called a $\mathbb{Z}$-basis or integral basis of $\mathcal{O}_K$); for instance, $(1, x, \ldots, x^{d-1})$ is such a basis if $\mathcal{O}_K = \mathbb{Z}[x]$ for some $x \in K$. The ring of integers $\mathcal{O}_K$ together with the embeddings $\sigma_1, \ldots, \sigma_d$ allow us to define the *discriminant* $\Delta_K$ of the number field $K$, which is given by $\Delta_K := \det((\sigma_i(x_j))_{i,j})^2$. Finally, we denote by $\mu_K$ the number of roots of unity in $K$.

**Cyclotomic Fields.** The main class of number fields we consider in this work is the class of *cyclotomic fields* [.py], which are number fields of the form $K = \mathbb{Q}(\omega_c)$ for $c \in \mathbb{N}$, where $\omega_c$ is a primitive $c$-th root of unity (i.e., $\omega_c \in \mathbb{C}$ and satisfies $\omega_c^c = 1$ and $\omega_c^j \neq 1$ for all $1 \leqslant j < c$). The *conductor* of $K$ is the minimal $c \in \mathbb{N}$ such that $K = \mathbb{Q}(\omega_c)$. If $c \in \mathbb{N}$ is odd, then $\mathbb{Q}(\omega_c) \cong \mathbb{Q}(\omega_{2c})$, and this criterion captures all isomorphic cyclotomic fields: the conductor $c$ of a cyclotomic field $K$ is either odd (in which case $K \cong \mathbb{Q}(\omega_{2c})$) or a multiple of 4. The number of roots of unity in a cyclotomic field $K$ of conductor $c$ equals $\mu_K = c$ when $c$ is even, and $\mu_K = 2c$ when $c$ is odd.

A cyclotomic field $K = \mathbb{Q}(\omega_c)$ satisfies $K \cong \mathbb{Q}[X]/\Phi_c(X)$ for the $c$-th cyclotomic polynomial $\Phi_c$. The degree of $K = \mathbb{Q}(\omega_c)$ equals $\phi(c)$, and its ring of integers is $\mathbb{Z}[\omega_c]$, which is isomorphic to $\mathbb{Z}[X]/\Phi_c(X)$. When $c > 2$, all of $K$'s embeddings are complex: $d_{\mathbb{R}} = 0$ and $d_{\mathbb{C}} = d/2$.

**Ideals, Modules, and Module Lattices.** A *fractional ideal* of $\mathcal{O}_K$ is an $\mathcal{O}_K$-submodule $\mathfrak{I} \subseteq K$ (i.e., it is closed under addition and under multiplication by elements of $\mathcal{O}_K$) for which there exists $x \in K\backslash\{0\}$ satisfying $x\mathfrak{I} \subseteq \mathcal{O}_K$. The algebraic norm $\mathrm{N}(\mathfrak{I})$ of a fractional ideal $\mathfrak{I}$ is defined as $\mathrm{N}(\mathfrak{I}) := [\mathcal{O}_K : x\mathfrak{I}]/|\mathrm{N}(x)|$ for any $x \in K\backslash\{0\}$ satisfying $x\mathfrak{I} \subseteq \mathcal{O}_K$.

An $\mathcal{O}_K$*-module* (or *module*) is a set of the form $\mathcal{M} = \sum_{i=1}^{r} \mathbf{b}_i \mathfrak{I}_i$ for nonzero fractional $\mathcal{O}_K$-ideals $\mathfrak{I}_1, \ldots, \mathfrak{I}_r$ and $K_{\mathbb{R}}$-linearly independent vectors $\mathbf{b}_1, \ldots, \mathbf{b}_r \in K_{\mathbb{R}}^m$ (for some $m > 0$). We say that $\mathcal{M}$ is a *free* module if the $\mathfrak{I}_i$'s are all equal to $\mathcal{O}_K$. We refer to the set $(\mathbf{b}_i, \mathfrak{I}_i)_{i=1}^r$ as a *pseudobasis* for $\mathcal{M}$, and remark that another way to represent module lattices is using module filtrations (see [MS20]). Such a pseudobasis is said to be *unital* if $1 \in \mathfrak{I}_i$ for all $i$ (equivalently, if $\mathcal{O}_K \subseteq \mathfrak{I}_i$ for all $i$).[8] We remark that $\mathcal{O}_K \subseteq \mathfrak{I}_i$ implies that $\mathrm{N}(\mathfrak{I}_i) \leqslant 1$.

The rank of $\mathcal{M}$ is defined to be $\mathrm{rank}_{K_{\mathbb{R}}}(\mathcal{M}) := \dim_{K_{\mathbb{R}}} \mathrm{span}_{K_{\mathbb{R}}}(\mathcal{M})$. For simplicity, we take $m = r = \mathrm{rank}_{K_{\mathbb{R}}}(\mathcal{M})$ in the remainder of this work. Denoting by $\mathbf{B}$ the matrix with columns $\mathbf{b}_1, \ldots, \mathbf{b}_r$, the determinant of $\mathcal{M}$ in $K_{\mathbb{R}}$ is defined as $\det_{K_{\mathbb{R}}}(\mathcal{M}) = \det(\overline{\mathbf{B}}^T \mathbf{B})^{1/2} \prod_{i=1}^{r} \mathfrak{I}_i$.

Through the embeddings of $K$, we can view $\mathcal{M}$ as an $rd$-dimensional Euclidean lattice. More precisely, writing $\boldsymbol{\sigma} \colon K \to \mathbb{C}^d$ for the *canonical embedding* of $K$, defined by $\boldsymbol{\sigma}(x) = (\sigma(x))_{\sigma \in \mathcal{E}}$, each $(x_1, \ldots, x_r) \in \mathcal{M} \subseteq K_{\mathbb{R}}^r$ gets mapped to a vector $(\boldsymbol{\sigma}(x_1), \ldots, \boldsymbol{\sigma}(x_r)) \in (\mathbb{R}^{d_{\mathbb{R}}} \times \mathbb{C}^{2d_{\mathbb{C}}})^r$. Since the complex embeddings come in conjugate pairs and $\mathbb{C} \cong \mathbb{R}^2$, the canonical embedding allows us to map $\mathcal{M}$ to an $rd$-dimensional Euclidean lattice in $\mathbb{R}^{rd}$ that preserves the geometry: the trace inner product of the module vectors in $K_{\mathbb{R}}^r$ is exactly equal to the Euclidean inner product of the corresponding embedded vectors. (For example, taking $K = \mathbb{Q}$ and thus $\mathcal{O}_K = \mathbb{Z}$, the canonical embedding is trivial and we recover the usual notion of Euclidean lattices in $\mathbb{Q}^r$.) It can be shown that the determinant of the module lattice $\mathcal{M}$ equals $\det_{\mathbb{Q}}(\mathcal{M}) := |\Delta_K|^{r/2} \mathrm{N}(\det_{K_{\mathbb{R}}}(\mathcal{M}))$.[9]

[8] Any pseudobasis can be turned into a unital pseudobasis using [LPSW19, Alg. 3.2].
[9] Sometimes a different normalization of embeddings is used (e.g., [PM19]), resulting in an additional factor of $2^{-rd_{\mathbb{C}}}$.

### 2.3 GSO over $K_\mathbb{R}$

Two vectors are said to be $K_\mathbb{R}$-linearly independent if and only if the zero vector cannot be expressed as a non-trivial $K_\mathbb{R}$-linear combination of them. This notion allows us to extend Gram-Schmidt orthogonalization to matrices $\mathbf{B} \in K_\mathbb{R}^{m\times r}$ with $K_\mathbb{R}$-linearly independent columns. (For instance, see [FS10, LPSW19].)

**Definition 2 (GSO over $K_\mathbb{R}$).** *Given $\mathbf{B} = (\mathbf{b}_1, \dots, \mathbf{b}_r)$ such that the $\mathbf{b}_i$ are $K_\mathbb{R}$-linearly independent, we define its GSO as $\mathbf{B}^* = (\mathbf{b}_1^*, \dots, \mathbf{b}_r^*)$, where $\mathbf{b}_1^* = \mathbf{b}_1$ and, for all $1 < i \leqslant r$,*

$$\mathbf{b}_i^* := \mathbf{b}_i - \sum_{j=1}^{i-1} \mu_{i,j}\mathbf{b}_j^* \quad \textit{with } \mu_{i,j} = \frac{\langle \mathbf{b}_i, \mathbf{b}_j^* \rangle_K}{\langle \mathbf{b}_j^*, \mathbf{b}_j^* \rangle_K} \textit{ for all } 1 \leqslant j < i.$$

More generally, we define the projection $\pi_i \colon \mathbf{x} \mapsto \mathbf{x} - \sum_{j=1}^{i-1} \frac{\langle \mathbf{x}, \mathbf{b}_j^* \rangle_K}{\langle \mathbf{b}_j^*, \mathbf{b}_j^* \rangle_K} \mathbf{b}_j^*$ for all $i$, so $\mathbf{b}_i^* = \pi_i(\mathbf{b}_i)$. Note that $\langle \mathbf{b}_i^*, \mathbf{b}_j^* \rangle_K = 0$ for all $i \neq j$.

Given a rank-$r$ module lattice $\mathcal{M}$ with pseudobasis $\mathfrak{B} = ((\mathbf{b}_i, \mathfrak{I}_i))_{i=1}^r$, we define the *projected module lattices* $\mathcal{M}(\mathfrak{B}_{[\![i;j]\!]}) = \pi_i(\mathbf{b}_i)\mathfrak{I}_i + \ldots + \pi_i(\mathbf{b}_j)\mathfrak{I}_j$ for $1 \leqslant i \leqslant j \leqslant r$. Note that $\mathcal{M}(\mathfrak{B}_{[\![i;j]\!]})$ has rank $j - i + 1$, and depends on $\mathfrak{B}$.

The determinant of $\mathcal{M}$ and of its projected module lattices can be expressed in terms of the GSO of the pseudobasis. Namely, for all $1 \leqslant i \leqslant j \leqslant r$, we have:

$$\det_\mathbb{Q}(\mathcal{M}(\mathfrak{B}_{[\![i;j]\!]})) = |\Delta_K|^{(j-i+1)/2} \cdot \prod_{k=i}^{j} \mathrm{N}(\mathbf{b}_k^*)\mathrm{N}(\mathfrak{I}_k).$$

## 3 Module-BKZ

After reviewing BKZ for arbitrary lattices and its corresponding slope prediction, we describe the modified BKZ algorithm for module lattices, *module-BKZ*, and present the specifics of our implementation.

### 3.1 BKZ and Corresponding Slope Prediction

Given a basis of an $n$-dimensional (Euclidean) lattice with GSO $(\mathbf{b}_1^*, \dots, \mathbf{b}_n^*)$, we define its $\mathbb{Q}$-*profile*[10] as the sequence $(\ell_1^\mathbb{Q}, \dots, \ell_n^\mathbb{Q})$, where $\ell_i^\mathbb{Q} := \ln\|\mathbf{b}_i^*\|$. As aforementioned, the more balanced the profile is, the better.

Turning a basis into a basis of better quality is the task of lattice reduction algorithms, such as the BKZ algorithm [Sch87, SE94]. In short, for blocksize $\beta$, the

[10] This is usually simply called the (log-)profile, but as we will generalize this notion to other number fields we explicit the number field for unambiguous discussions.

BKZ algorithm aims to return a lattice basis $\mathbf{B}$ such that its GSO $(\mathbf{b}_1^*, \ldots, \mathbf{b}_n^*)$ is BKZ$^\beta$*-reduced*, i.e., it satisfies

$$\|\mathbf{b}_i^*\| = \lambda_1(\mathcal{L}(\mathbf{B}_{[\![i;\min(i+\beta-1,n)]\!]})) \qquad \forall\, 1 \leqslant i \leqslant n,$$

where $\mathcal{L}(\mathbf{B}_{[\![i;j]\!]})$ is the lattice generated by $(\pi_i(\mathbf{b}_i), \ldots, \pi_i(\mathbf{b}_j))$ and $\pi_i$ is the linear projection orthogonal to the span of $\mathbf{b}_1, \ldots, \mathbf{b}_{i-1}$. Approximation variants of BKZ exist as well, but are beyond the scope of this paper.

**Slope Prediction under the Geometric Series Assumption.** One can predict the slope of the $\mathbb{Q}$-profile of a BKZ$^\beta$-reduced basis. Such predictions are often based on the empirical observation that the $\mathbb{Q}$-profile of a BKZ$^\beta$-reduced basis tends to resemble a descending straight line for sufficiently large $\beta \ll n$ (say $\beta \geqslant 50$). This is formalized by the Geometric Series Assumption (GSA), as first proposed by Schnorr [Sch03].

**Heuristic 1 (Euclidean-Lattice GSA).** *Let $\beta \ll n$ be sufficiently large. There is a constant $\alpha_\mathbb{Q} > 1$ (depending only on $\beta$) such that the $\mathbb{Q}$-profile of a* BKZ$^\beta$*-reduced basis of a random $n$-dimensional Euclidean lattice of fixed determinant satisfies:*

$$\mathbb{E}[\ell_i^\mathbb{Q}] = \mathbb{E}[\ell_1^\mathbb{Q}] - (i-1)\ln\alpha_\mathbb{Q} \qquad \forall\, 1 \leqslant i \leqslant n. \tag{GSA}$$

Due to the lattice invariant $\ln\det_\mathbb{Q}(\mathcal{L}) = \sum_{i=1}^n \ell_i^\mathbb{Q}$ and the definition of BKZ reduction, the GSA is equivalent to the following prediction of the *expected* $\mathbb{Q}$-*slope*, defined as $\mathrm{slope}_\mathbb{Q}(\mathrm{BKZ}^\beta) := -\ln\alpha_\mathbb{Q}$. (See, for example, [AD21, dBvW25].)

**Heuristic Claim 1 ($\mathbb{Q}$-Profile of BKZ$^\beta$ under GSA).** *Let $\mathcal{L}$ be a random $n$-dimensional Euclidean lattice of fixed determinant, and let $\mathbf{B}$ be a* BKZ$^\beta$*-reduced basis of $\mathcal{L}$ for some sufficiently large $\beta \ll n$. Then the GSA predicts*

$$\mathbb{E}[\ell_i^\mathbb{Q}] = \frac{n+1-2i}{2}\ln\alpha_\mathbb{Q} + \frac{1}{n}\ln\det\nolimits_\mathbb{Q}(\mathcal{L})$$

*for all $1 \leqslant i \leqslant n$, where:*

$$\ln\alpha_\mathbb{Q} = \frac{2}{\beta-1}\,\mathbb{E}_\mathbf{s}[\ln\|\mathbf{s}\|]$$

*where $\mathbf{s}$ is a shortest vector in one of the random normalized projected lattices $\mathcal{L}(\mathbf{B}_{[\![j;j+\beta-1]\!]})^{(1)}$ for $1 \leqslant j < n-\beta$.*

**Further Prediction using the Gaussian Heuristic.** In the analysis of BKZ algorithms, it is standard to go one step further in the slope prediction by assuming that the normalized projected lattices $\mathcal{L}(\mathbf{B}_{[\![j;j+\beta-1]\!]})^{(1)}$ behave as random $\beta$-dimensional lattices of unit determinant, allowing to invoke the Gaussian Heuristic to estimate $\mathbb{E}_\mathbf{s}[\ln\|\mathbf{s}\|]$.

More precisely, the Gaussian Heuristic states that, for an $n$-dimensional Euclidean lattice $\mathcal{L}$ and a measurable set $\mathcal{B} \subseteq \mathrm{span}(\mathcal{L})$, the number of nonzero lattice vectors in $\mathcal{L} \cap \mathcal{B}$ approximately equals $\mathrm{vol}(\mathcal{B})/\det_{\mathbb{Q}}(\mathcal{L})$. Applying this heuristic to the $n$-dimensional Euclidean unit ball $\mathcal{B}_n$ gives a prediction of $\lambda_1(\mathcal{L}) \approx (\det_{\mathbb{Q}}(\mathcal{L})/\mathrm{vol}(\mathcal{B}_n))^{1/n}$ for the first minimum of $\mathcal{L}$.

For a well-defined notion of *random* lattices of unit determinant, there exists a more precise and formal result regarding the expectation of $\lambda_1(\mathcal{L})$, namely $\lambda_1(\mathcal{L})^n \cdot \mathrm{vol}(\mathcal{B}_n)$ converges in distribution to the exponential distribution of parameter 1/2 [BSW18, Theorem 1] (attributed to [Söd11, Che13]). As we study the (log-)profile, we are more interested in $\mathbb{E}[\ln \lambda_1(\mathcal{L})]$, but this value can also be predicted by the aforementioned theorem. Indeed, the logarithm of an exponential distribution can be expressed in terms of a Gumbel distribution, whose mean is known. Note that this formal result still downgrades to the following *heuristic* when applied to the lattices appearing in BKZ, as their distributions are not yet well understood.

**Heuristic 2** **($\ln \lambda_1$ under the Euclidean-Lattice Gaussian Heuristic [.py]).** *Let $\mathcal{L}$ be a random $n$-dimensional lattice of unit determinant. Its expected logarithmic first minimum under the Gaussian Heuristic is given by*

$$\mathbb{E}[\ln \lambda_1(\mathcal{L})] = \mathrm{lgh}_{\mathbb{Q}}(n) \tag{GH}$$

*where* $\mathrm{lgh}_{\mathbb{Q}}(n) := \frac{1}{n}\left(\ln(2/\mathrm{vol}(\mathcal{B}_n)) - \gamma\right)$, *with* $\gamma \approx 0.57721$ *denoting the Euler-Mascheroni constant.*

It is therefore common to heuristically predict the slope of the $\mathbb{Q}$-profile as

$$\mathrm{slope}_{\mathbb{Q}}(\mathrm{BKZ}^{\beta}) = -\frac{2}{\beta - 1}\,\mathrm{lgh}_{\mathbb{Q}}(\beta).$$

*Remark 3 (Gaussian Heuristic for Lattices of Arbitrary Determinants).* This heuristic extends to lattices of arbitrary determinant after appropriate scaling. Namely, given a random $n$-dimensional lattice $\mathcal{L}$ of fixed determinant $D$, the Gaussian Heuristic predicts: $\mathbb{E}[\ln \lambda_1(\mathcal{L})] = \mathrm{lgh}_{\mathbb{Q}}(n) + \frac{1}{n}\ln D$.

### 3.2 Module-BKZ

The BKZ algorithm merely treats a rank-$r$ module $\mathcal{M} \subseteq K_{\mathbb{R}}^r$ as an $rd$-dimensional Euclidean lattice, where $d = \deg(K)$, thereby ignoring the underlying module structure. Instead, we consider a generalization of BKZ, which we call $\mathrm{mBKZ}_K^{\beta_K}$ for a parameter $2 \leqslant \beta_K \leqslant r$. This module-BKZ algorithm is presented in Algorithm 1 and finds short vectors in the rank-$r$ module using an SVP oracle on rank-$\beta_K$ projected modules, thereby generalizing module-LLL [LPSW19] to $\beta_K \geqslant 2$. These projected modules correspond to Euclidean lattices of dimension $\beta_K d$, which makes it natural to compare $\mathrm{mBKZ}_K^{\beta_K}$ to $\mathrm{BKZ}^{\beta_K d}$. The output of Algorithm 1 is an $\mathrm{mBKZ}_K^{\beta_K}$-reduced pseudobasis of $\mathcal{M}$, defined as follows.

**Definition 4** **($\mathrm{mBKZ}_K^{\beta_K}$-Reduced).** *For $2 \leqslant \beta_K \leqslant r$, we say that a pseudobasis $\mathfrak{B}$ of a rank-$r$ module $\mathcal{M}$ is $\mathrm{mBKZ}_K^{\beta_K}$-reduced if*

$$\|\mathbf{b}_i^*\| = \lambda_1(\mathcal{M}(\mathfrak{B}_{\llbracket i;\min(i+\beta_K-1,r)\rrbracket})) \qquad \forall\, 1 \leqslant i \leqslant r.$$

**Algorithm 1:** Module-BKZ algorithm (high-level)

**Input:** Unital pseudobasis $\mathfrak{B}$ of a rank-$r$ module $\mathcal{M}$; parameter $\beta_K$
**Output:** Unital $\mathrm{mBKZ}_K^{\beta_K}$-reduced pseudobasis

```
1 while 𝔅 is not mBKZ_K^{β_K}-reduced do
2     for i = 1, ..., r do
3         Let v = SVP(M(𝔅_⟦i;min(i+β_K−1,r)⟧))
4         Let ℑ be such that vℑ = vK ∩ M(𝔅_⟦i;min(i+β_K−1,r)⟧)
5         Lift vℑ to wℑ ⊆ M satisfying π_i(w) = v
6         Insert (w, ℑ) into 𝔅 at position i and remove linear dependencies
7 return 𝔅
```

We refer to a single execution of the while-loop as an *mBKZ tour*. Note that Line 5 is essentially Babai lifting [Bab86], and Line 6 can be achieved using (e.g.) the algorithm from [FS10, Theorem 4]. While Algorithm 1 may encounter non-unital pseudobases during an mBKZ tour, the final output is always unital by construction. We remark that formal variants of LLL and BKZ typically consider an *approximate* SVP oracle to enforce significant improvement at each insertion step. As we are not concerned with guaranteed termination, our model instead uses an exact SVP oracle in Line 3. This approach assumes approximate convergence after sufficiently many mBKZ tours, which is at least heuristically supported by [HPS11].

*Remark 5.* Note that the fractional ideal $\mathfrak{I}$ considered in Line 4 contains $\mathcal{O}_K$ and is the ideal of minimal norm that satisfies $\mathbf{v}\mathfrak{I} \subseteq \mathcal{M}(\mathfrak{B}_{\llbracket i;\min(i+\beta_K-1,n)\rrbracket})$. Indeed, consider a vector $\mathbf{v} \in \mathcal{M}'$ for some module $\mathcal{M}'$, and let $\mathfrak{I}$ be the fractional ideal such that $\mathbf{v}\mathfrak{I} = \mathbf{v}K \cap \mathcal{M}'$. Then $\mathfrak{I} = \{x \in K : x\mathbf{v} \in \mathcal{M}'\}$. Since $x\mathbf{v} \in \mathcal{M}'$ for all $x \in \mathcal{O}_K$, we have that $\mathcal{O}_K \subseteq \mathfrak{I}$, so $\mathcal{N}(\mathfrak{I}) \leqslant 1$. Moreover, for all fractional ideals $\mathfrak{I}'$ such that $\mathbf{v}\mathfrak{I}' \subseteq \mathcal{M}'$, we have $\mathfrak{I}' \subseteq \mathfrak{I}$, and thus $\mathcal{N}(\mathfrak{I}) \leqslant \mathcal{N}(\mathfrak{I}')$.

### 3.3 Implementation

For our experiments, we developed a `Python` implementation of module-BKZ for cyclotomic fields, which is publicly available, together with the experiment and prediction scripts, and the generated data. Instead of redeveloping the entire lattice-reduction stack to obtain one specialized to module lattices, we rely on existing libraries `fplll`/`fpylll` [dt23] and `G6K` [ADH+19]. When implementing the SVP oracle from G6K [.py], we choose less aggressive strategies than typically done to be almost certain to really solve SVP, as our purpose is more to understand and predict than to fine-tune for speed.

The module lattices considered in our experiments are natural $\mathcal{O}_K$-analogs of $q$-ary lattices, namely lattices defined by random linear equations modulo $q\mathcal{O}_K$ [.py], though we remark that our implementation is not restricted to those.

The choice of using existing lattice-reduction libraries raises two technical difficulties: protecting the module structure from the whims of libraries that are designed to only consider $\mathbb{Z}$-bases, and having a geometrically faithful embedding of the cyclotomic rings that remains integral. We discuss our solutions to both issues.

*Restructured $\mathbb{Z}$-bases.* Our objective is to represent a module lattice $\mathcal{M}$ by a $\mathbb{Z}$-basis, while maintaining enough of its structure. Let $(\mathbf{b}_1, \ldots, \mathbf{b}_{rd}) \in K_{\mathbb{R}}^{r \times rd}$ be a $\mathbb{Z}$-basis of $\mathcal{M}$ at some stage of the algorithm. Naively, one would be tempted to enforce that, for all blocks $i$, the corresponding vectors $\mathbf{b}_{(i-1)d+1}, \ldots, \mathbf{b}_{(i-1)d+d}$ $\mathbb{Z}$-generate a rank-1 module. This constraint may be broken when applying size reduction, the most frequent basis maintenance task of the lattice-reduction library. Size reduction can hardly be avoided for reasons of numerical stability, and replacing it by a module-lattice-analog of size reduction would require significant modifications in `fplll`.

Instead, it suffices to impose the constraint that $(\mathbf{b}_1, \ldots, \mathbf{b}_d)$ is indeed the $\mathbb{Z}$-basis of a rank-1 module [.py], and that this property holds recursively for the rest of the basis (i.e., $\mathbf{b}_{d+1}, \ldots, \mathbf{b}_{rd}$) projected orthogonally to $\mathbf{b}_1, \ldots, \mathbf{b}_d$. This property is not affected by size reduction, and is sufficient to implement module-BKZ.

This structure is going to be broken whenever we solve SVP on a block by `G6K`. We now want a procedure to repair the module structure; more precisely, given a certain first vector $\mathbf{b}_1$, we would like to enforce that $(\mathbf{b}_1, \ldots, \mathbf{b}_d)$ is a $\mathbb{Z}$-basis of $\mathbf{b}_1 K \cap \mathcal{M}$ [.py], which is a rank-1 module containing $\mathbf{b}_1 \mathcal{O}_K$. Hence, we first insert a $\mathbb{Z}$-basis of $\mathbf{b}_1 \mathcal{O}_K$ into the current $\mathbb{Z}$-basis of $\mathcal{M}$, and would then like to run LLL to eliminate linear dependencies [.py].

Yet, LLL may find shorter vectors for this block and thus re-break the module structure of the first block: the first $d$ vectors may not $\mathbb{Z}$-generate $\mathbf{b}_1 K \cap \mathcal{M}$. In principle, one can prevent LLL from doing just that while still eliminating linear dependencies, simply by running LLL (in Line 6) with parameter $\delta$ very close to 0, so that vectors are less likely to be swapped.

Unfortunately, the `fplll` API forbids setting $\delta \leqslant 1/4$; the reason might be that there is no absolute guarantee in terms of basis quality for such small $\delta$. Yet, for such parameters LLL still guarantees that the potential does not increase and that linear dependencies are eliminated. We argue that it would be desirable to relax the API.

Fortunately, we do not have to resort to patching `fplll` to get our prototype running, at least for some cyclotomic rings of interest (say with conductor $\leqslant$ 16); we found out experimentally that repeated attempts to restructure, with decreasing $\delta$ from 0.99 would eventually succeed before reaching 1/4 [.py]. Running

BKZ progressively with increasing blocksizes also seems to help avoiding the issue.

*Cyclic Embedding.* A second technical difficulty with our approach comes from the definition of inner products. Our implementation considers module lattices over $K = \mathbb{Q}(\omega_c)$, so we are working in the cyclotomic ring $\mathcal{O}_K = \mathbb{Z}[\omega_c]$ with the trace inner product. However, the `fplll` library requires the input lattice to be represented with integer coefficients, and considers the standard inner product over $\mathbb{R}^n$. If we represent $\mathbb{Z}[\omega_c] \cong \mathbb{Z}[X]/\Phi_c(X)$ using the coefficients of a polynomial $\sum_{i=0}^{c-1} a_i X^i$, then the inner product does not correspond to the desired trace inner product. We could represent a vector by its embeddings to have the correct inner product, but those values are defined over $\mathbb{C}$ and thus may be irrational. The following *cyclic embedding* solves this conundrum; an idea that already underpins several existing works [DD12, BDF18, DP16].

Consider the cyclic ring $\mathcal{C}_c = \mathbb{R}[X]/(X^c - 1)$, and view it as a Euclidean vector space in the standard way, meaning that elements of $\mathcal{C}_c$ are polynomials $\sum_{i=0}^{c-1} a_i X^i$ represented by their coefficients $a_i$, naturally leading to the *coefficient inner product* $\langle \sum_{i=0}^{c-1} a_i X^i, \sum_{i=0}^{c-1} b_i X^i \rangle_{\mathrm{coef}} := \sum_{i=0}^{c-1} a_i b_i$. The discrete Fourier transform $F_c$ sends such a polynomial $P(X) = \sum_{i=0}^{c-1} a_i X^i$ to the vector $(P(\omega_c^j))_{j=0}^{c-1}$ in $\mathbb{C}^c$. It is well known that its scaling $\frac{1}{\sqrt{c}} F_c$ is an isometry.

Similarly, the trace inner product of $\mathbb{Q}(\omega_c)$ is isometric to the standard inner product of $\mathbb{C}^{\phi(n)}$ via the canonical embedding. The latter Hermitian space is trivially embedded into $\mathbb{C}^c$ by padding $c - \phi(c)$ many zero coordinates at positions corresponding to the non-primitive $c$-th roots of unity. This allows us to define the cyclic embedding by completing a commutative diagram, as depicted in Figure 3 [.py].

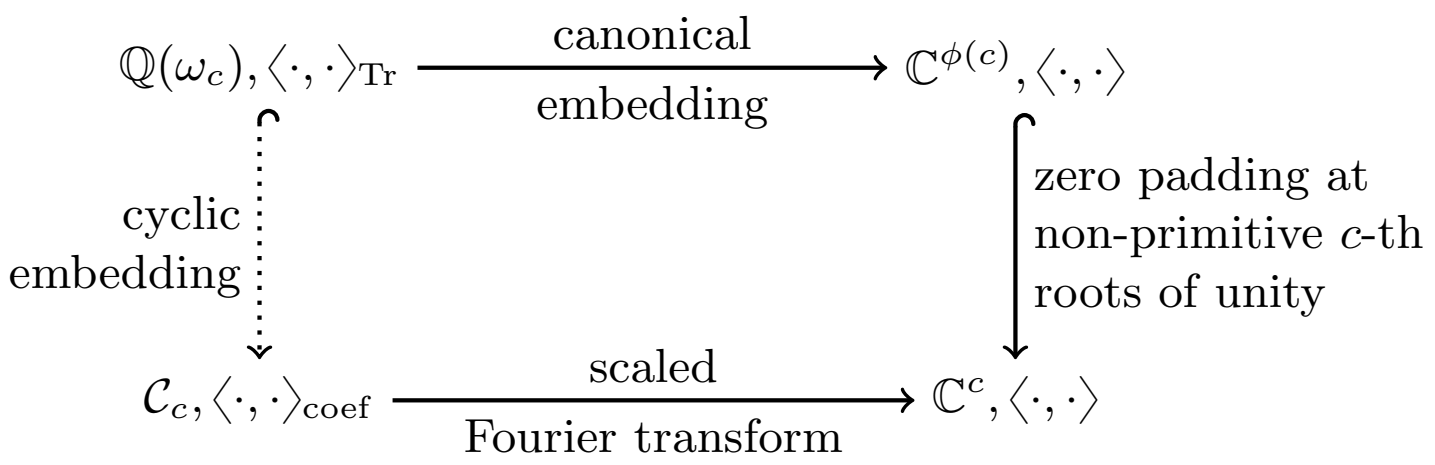

Fig. 3: Commutative diagram defining the cyclic embedding.

One further notices that the ring of integers $\mathbb{Z}[\omega_c] \subseteq \mathbb{Q}(\omega_c)$ is represented by elements of $\frac{1}{\sqrt{c}}\mathbb{Z}[X]/(X^c - 1) \subseteq \mathcal{C}_c$ under this isometric map. Indeed, the cyclic embedding of $x \in \mathbb{Z}[\omega_c]$ is the polynomial $\sum_{i=0}^{c-1} a_i X^i \in \mathcal{C}_c$ with coefficients

$$a_i = \frac{1}{\sqrt{c}} \sum_{j=1}^{d} \sigma_j(x)\sigma_j(\omega_c^{-i}) = \frac{1}{\sqrt{c}} \sum_{j=1}^{d} \sigma_j(x\omega_c^{-i}) = \frac{1}{\sqrt{c}} \mathrm{Tr}(x\omega_c^{-i}).$$

As both $x$ and $\omega_c$ lie in $\mathbb{Z}[\omega_c]$, we have $x\omega_c^{-i} \in \mathbb{Z}[\omega_c]$, so its trace lies in $\mathbb{Z}$. Hence, multiplying the cyclic embeddings of elements of $\mathbb{Z}[\omega_c]$ by a factor of $\sqrt{c}$ results in a scaled isometric embedding of $\mathbb{Z}[\omega_c]$ into $\mathbb{Z}^c$, a space that is compatible with libraries such as `fplll`. Note that the embedding dimension $c$ is strictly larger than the degree $\phi(c)$ of the number field, but this is not an issue as the aforementioned lattice-reduction library is perfectly capable of working with lattices embedded in a vector space of larger dimension.

# 4 Prediction of the Module-BKZ Profile

In this section, we present a slope prediction for (structured) module-BKZ, motivated by numerical observations and supported by theoretical analysis, which makes use of a module-lattice analog of the Geometric Series Assumption (GSA).

Let $K$ be a number field of degree $d$ and discriminant $\Delta_K$. Consider a random rank-$r$ module $\mathcal{M}$ over $K$ of fixed determinant, and a sufficiently large $\beta_K$ satisfying $2 \leqslant \beta_K \ll r$. Viewing $\mathcal{M}$ as an $rd$-dimensional Euclidean lattice $\mathcal{L}$, the analysis from Section 3.1 (using the GSA) implies that the unstructured $\mathrm{BKZ}^{\beta_K d}$ algorithm results in a slope prediction of

$$\mathrm{slope}_{\mathbb{Q}}(\mathrm{BKZ}^{\beta_K d}) = -\frac{2}{\beta_K d - 1} \mathbb{E}_{\mathbf{s}}[\ln\|\mathbf{s}\|] \tag{2}$$

where the random variable $\mathbf{s} \in \mathbb{R}^{\beta_K d}$ reflects the behavior of BKZ. Specifically, $\mathbf{s}$ is a shortest vector in a random $\beta_K d$-dimensional lattice of unit determinant. We recall from Section 3.1 that by invoking the Gaussian Heuristic one can predict $\mathbb{E}_s[\ln\|\mathbf{s}\|]$ as $\mathrm{lgh}_{\mathbb{Q}}(\beta_K d)$, as is common in BKZ-slope analysis.

In Section 4.1, we use a module-lattice analog of the GSA to derive the following slope prediction [.py] for the $\mathrm{mBKZ}_K^{\beta_K}$ algorithm:

$$\mathrm{slope}_{\mathbb{Q}}(\mathrm{mBKZ}_K^{\beta_K}) = -\frac{2}{\beta_K d - d}\Bigg( \underbrace{\mathbb{E}_{\mathbf{s}}[\ln\|\mathbf{s}\|]}_{t_1} + \underbrace{\frac{1}{2d}\ln\frac{|\Delta_K|}{d^d}}_{t_2} + \underbrace{\mathbb{E}_{\mathbf{s}}\left[\ln\frac{\sqrt{d}\mathrm{N}(\mathbf{s})^{1/d}}{\|\mathbf{s}\|}\right]}_{t_3} + \underbrace{\frac{1}{d}\mathbb{E}_{\mathfrak{I}}[\ln \mathrm{N}(\mathfrak{I})]}_{t_4} \Bigg) \tag{3}$$

where the random variables $\mathbf{s} \in K_{\mathbb{R}}^{\beta_K}$ and $\mathfrak{I} \supseteq \mathcal{O}_K$ reflect the behavior of module-BKZ. Specifically, $\mathbf{s}$ is a shortest vector in a random rank-$\beta_K$ module lattice of unit determinant, and $\mathfrak{I}$ is a fractional ideal in the random mBKZ-reduced pseudobasis.[11]

Apart from the terms $t_1, t_2, t_3, t_4$, the slope predictions of unstructured BKZ in Equation (2) and structured module-BKZ in Equation (3) also differ in the denominator, changing from $\beta_K d - 1$ to $\beta_K d - d$.

After deriving this slope prediction, we dive into the four main terms $t_1, t_2, t_3, t_4$, enabling us to compare with the usual BKZ slope prediction from Equation (2). We refer to these four terms as:

$t_1$: the module-lattice analog of $\mathrm{lgh}_{\mathbb{Q}}(\beta_K d)$,

$t_2$: the discriminant gap,

$t_3$: the skewness gap,

$t_4$: the index gap.

Note that $t_2$ is fully determined by $K$. The other terms have an upper bound independent of the pseudobasis: for $t_1$ there is Minkowski's bound, while $t_3 \leqslant 0$ by the arithmetic-geometric inequality, and $t_4 \leqslant 0$ since the pseudobasis is unital. Sections 4.2 up to 4.5 are aimed at providing an estimate for those terms, rather than just a generic bound.

In Section 4.6, we use our analysis of the four terms in order to conclude with an upper and lower bound on the module-BKZ $\mathbb{Q}$-slope prediction (Equation (3)), yielding the prediction interval shown in Figure 1.

*Remark 6 (Terminology).* As $t_3 = 0$ if and only if $|\sigma(\mathbf{s})|^2$ has the same value for all embeddings $\sigma$, $t_3$ measures how *skewed* these values are. Moreover, $t_4$ is defined by the norm of ideals $\mathfrak{I} \supseteq \mathcal{O}_K$, which satisfy $\mathrm{N}(\mathfrak{I}) = 1/\mathrm{N}(\mathfrak{I}^{-1})$ where $\mathrm{N}(\mathfrak{I}^{-1})$ equals the *index* of $\mathfrak{I}^{-1}$ as a subgroup of $\mathcal{O}_K$.

## 4.1 Module-Lattice GSA and Corresponding Slope Prediction

Consider a pseudobasis $((\mathbf{b}_i, \mathfrak{I}_i))_{i=1}^r$ of a rank-$r$ module $\mathcal{M}$ over $K$. Let $\mathbf{z}_1^*, \ldots, \mathbf{z}_{rd}^*$ denote the GSO of a $\mathbb{Z}$-basis obtained by embedding $K$ into $\mathbb{R}^d$ and successively mapping each of the $r$ components of the pseudobasis. Recall from Section 3.1 that the corresponding $\mathbb{Q}$-*profile* is defined as the sequence $(\ell_i^{\mathbb{Q}})_{i=1}^{rd}$ where:

$$\ell_i^{\mathbb{Q}} := \ln\|\mathbf{z}_i^*\| \qquad \forall\, 1 \leqslant i \leqslant rd.$$

In addition, we define the corresponding $K$-*profile* as the sequence $(\ell_i^K)_{i=1}^r$ where:

$$\ell_i^K := \ln \det\nolimits_{\mathbb{Q}}(\mathbf{b}_i^* \cdot \mathfrak{I}_i) = \sum_{j=1}^{d} \ell_{(i-1)d+j}^{\mathbb{Q}} \qquad \forall\, 1 \leqslant i \leqslant r.$$

[11] As is standard in the analysis of BKZ algorithms, these distributions are not formally defined, which is why most of our analysis remains heuristic (recall Remark 1).

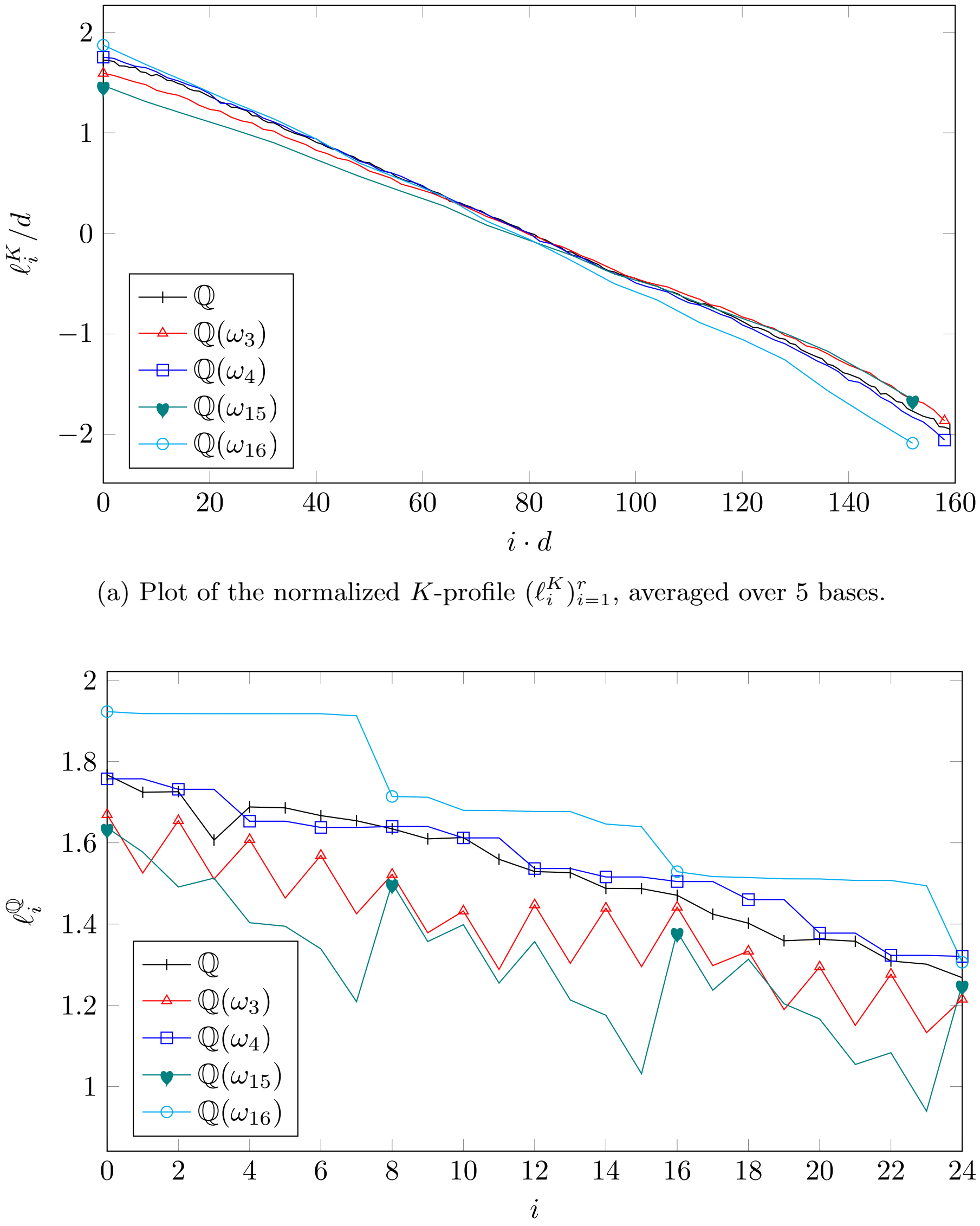

(a) Plot of the normalized $K$-profile $(\ell_i^K)_{i=1}^r$, averaged over 5 bases.

(b) Plot of $\mathbb{Q}$-profiles $(\ell_i^{\mathbb{Q}})_{i=1}^{rd}$, zoomed in.

Fig. 4: The profiles resulting from mBKZ$_K^{\beta_K}$ in dimension $rd = 160$ with blocksize $\beta_K d = 64$ after 30 tours, where $d = \deg(K)$ and $K = \mathbb{Q}(\omega_c)$ for $c = 1, 3, 4, 15, 16$ [.py]. The respective degrees are $d = 1, 2, 2, 8, 8$.

In particular, $\sum_{i=1}^r \ell_i^K = \ln \det_{\mathbb{Q}}(\mathcal{M})$, and each $\ell_i^K/d$ denotes the average log-norm of the GSO vectors corresponding to the ideal $\mathbf{b}_i \mathfrak{I}_i$.

**Geometric Series Assumption for Module Lattices.** Figure 4a illustrates the $K$-profile of mBKZ-reduced bases of module lattices over the cyclotomic fields $\mathbb{Q}(\omega_3)$, $\mathbb{Q}(\omega_4)$, $\mathbb{Q}(\omega_{15})$, and $\mathbb{Q}(\omega_{16})$. It is compared with $\mathbb{Q}$, the unstructured case. A first observation is that the GSA seems to generalize to the $K$-profile of mBKZ-reduced bases, up to a tail phenomenon that is well known for regular BKZ [AD21, Def. 9]. However, we notice that the module-BKZ slope varies: it is better (flatter) for $\mathbb{Q}(\omega_3)$ and $\mathbb{Q}(\omega_{15})$, and worse for $\mathbb{Q}(\omega_4)$ and $\mathbb{Q}(\omega_{16})$.[12]

Moreover, we observe various periodic patterns in the corresponding $\mathbb{Q}$-profiles shown in Figure 4b. These repeated patterns correspond to the profile of a (reduced) $\mathbb{Z}$-basis of $\mathcal{O}_K$, or a small distortion thereof. In the case of $\mathbb{Q}(\omega_4)$, $\mathcal{O}_K \cong \mathbb{Z}^2$ (up to scaling), and we have $\|\mathbf{b}_1\| = \|\mathbf{b}_2^*\| = 1$: the pattern is perfectly flat on each period. For $\mathbb{Q}(\omega_3)$, $\mathcal{O}_K$ is a scaled hexagonal lattice, with $\|\mathbf{b}_1\| = 1, \|\mathbf{b}_2^*\| = \frac{\sqrt{3}}{2}$. For $\mathbb{Q}(\omega_{16})$, we have $\mathcal{O}_K \cong \mathbb{Z}^8$, but the pattern is not exactly flat nor perfectly periodic: $\mathcal{O}_K$ seems to be slightly and randomly *skewed*.

These observations regarding Figure 4a motivate the following module-lattice analog of the GSA, as already proposed for module-LLL in [KK24, Heuristic 3].[13]

**Heuristic 3 (Module-Lattice GSA).** *Let $K$ be a number field and let $\beta_K \ll r$ be sufficiently large. There is a constant $\alpha_K > 1$ (depending on $\beta_K$) such that the $K$-profile of an $\mathrm{mBKZ}_K^{\beta_K}$-reduced pseudobasis of a random rank-$r$ module $\mathcal{M}$ over $K$ of fixed determinant satisfies:*

$$\mathbb{E}[\ell_i^K] = \mathbb{E}[\ell_1^K] - (i-1)\ln\alpha_K \qquad \forall\, 1 \leqslant i \leqslant r. \qquad \text{(mGSA)}$$

We recall that $\ell_i^K = \ln\det_{\mathbb{Q}}(\mathbf{b}_i^* \cdot \mathfrak{I}_i) = \frac{1}{2}\ln|\Delta_K| + \ln \mathrm{N}(\mathbf{b}_i^*)\mathrm{N}(\mathfrak{I}_i)$ for $1 \leqslant i \leqslant r$.

**Slope Prediction.** Similar to the unstructured case (Section 3.1), the module invariant $\det_{\mathbb{Q}}(\mathcal{M})$ and the definition of mBKZ reduction allow us to translate Heuristic 3 into the following prediction of the *expected $K$-slope*, defined as $\mathrm{slope}_K(\mathrm{mBKZ}_K^{\beta_K}) := -\ln\alpha_K$ [.py].

**Heuristic Claim 2 ($K$-Profile of $\mathrm{mBKZ}_K^{\beta_K}$ under mGSA).** *Let $K$ be a number field of degree $d$. Let $\mathcal{M}$ be a random rank-$r$ module over $K$ of fixed determinant, and let $\mathfrak{B}$ be an $\mathrm{mBKZ}_K^{\beta_K}$-reduced pseudobasis of $\mathcal{M}$ for some sufficiently large $\beta_K \ll r$. Then the module-lattice GSA predicts*

$$\mathbb{E}[\ell_i^K] = \frac{r+1-2i}{2}\ln\alpha_K + \frac{1}{r}\ln\det_{\mathbb{Q}}(\mathcal{M})$$

[12] This discrepancy can be explained by the difference in discriminant gap: cyclotomic fields $K$ with a power-of-two conductor $c$ have $|\Delta_K| = \phi(c)^{\phi(c)}$, whereas a strict inequality holds for conductors $c = 3$ and $c = 15$, contributing to a flatter profile. (See Section 4.3.)

[13] Technically, the heuristic in [KK24] considers module-LLL using an algebraic SVP oracle, minimizing the *algebraic* norm of the basis vectors, whereas our definition of module-BKZ deals with minimizing the Euclidean norm (recall Open Question 6).

*for all* $1 \leqslant i \leqslant r$*, where:*

$$\ln \alpha_K = \frac{2d}{\beta_K - 1}\Big(\mathbb{E}_{\mathbf{s}}[\ln\|\mathbf{s}\|] + \frac{1}{2d}\ln\frac{|\Delta_K|}{d^d} + \mathbb{E}_{\mathbf{s}}\Big[\ln\frac{\sqrt{d}\mathrm{N}(\mathbf{s})^{\frac{1}{d}}}{\|\mathbf{s}\|}\Big] + \frac{1}{d}\mathbb{E}_{\mathfrak{I}}[\ln \mathrm{N}(\mathfrak{I})]\Big) \tag{4}$$

*where* $\mathbf{s}$ *is a shortest vector in one of the random normalized projected module lattices* $\mathcal{M}(\mathfrak{B}_{[\![j;j+\beta_K-1]\!]})^{(1)}$ *for* $1 \leqslant j < r - \beta_K$*, and* $\mathfrak{I} \supseteq \mathcal{O}_K$ *is a fractional ideal in the random pseudobasis* $\mathfrak{B}$*.*

*Remark 7 (Prediction of the* $\mathbb{Q}$*-Slope).* While the Euclidean-lattice GSA (Heuristic 1) may not hold locally, it is globally plausible and allows us to translate the slope of the $K$-profile back to that of the corresponding $\mathbb{Q}$-profile. Namely, we predict:

$$\mathrm{slope}_{\mathbb{Q}}(\mathrm{mBKZ}_K^{\beta_K}) = \frac{1}{d^2}\,\mathrm{slope}_K(\mathrm{mBKZ}_K^{\beta_K}) = -\frac{1}{d^2}\ln\alpha_K$$

resulting in the slope prediction given in Equation (3). Note that the first equality holds under Heuristic 1: $\ln\alpha_K = \ell_1^K - \ell_2^K = \sum_{j=1}^{d}\ell_j^{\mathbb{Q}} - \sum_{j=1}^{d}\ell_{d+j}^{\mathbb{Q}} = d^2\ln\alpha_{\mathbb{Q}}$, where we use Equation (mGSA), the definition of the $\ell_i^K$'s, and Equation (GSA).

*Justification of Heuristic Claim 2.* Let $\mathfrak{B} := ((\mathbf{b}_i, \mathfrak{I}_i))_{i=1}^r$ be the random pseudobasis output by module-BKZ. We first show that Heuristic 3 implies Equation (4). Let $D_j := \det_{\mathbb{Q}}(\mathcal{M}(\mathfrak{B}_{[\![j;j+\beta_K-1]\!]}))$ for some $1 \leqslant j < r - \beta_K$. By definition of the $K$-profile and by applying Equation (mGSA) to all $i \in \{1,\ldots,\beta_K\}$, we have $\mathbb{E}[\ln(D_j)] = \sum_{i=0}^{\beta_K-1}\mathbb{E}[\ell_{j+i}^K] = \beta_K(\mathbb{E}[\ell_j^K] - \frac{\beta_K-1}{2}\ln\alpha_K)$, so we obtain $\ln\alpha_K = \frac{2}{\beta_K-1}(\mathbb{E}[\ell_j^K] - \frac{1}{\beta_K}\mathbb{E}[\ln(D_j)])$.

Moreover, $\ell_j^K = \frac{1}{2}\ln|\Delta_K| + \ln\mathrm{N}(\mathbf{b}_j) + \ln\mathrm{N}(\mathfrak{I}_j)$, which can be rewritten as $\ell_j^K = d\ln(\|\mathbf{b}_j\|) + \frac{1}{2}\ln\frac{|\Delta_K|}{d^d} + d\ln\frac{\sqrt{d}\mathrm{N}(\mathbf{b}_j)^{1/d}}{\|\mathbf{b}_j\|} + \ln\mathrm{N}(\mathfrak{I}_j)$. Hence, $\frac{1}{d}(\ell_j^K - \frac{1}{\beta_K}\ln(D_j)) = \ln\frac{\|\mathbf{b}_j\|}{D_j^{1/\beta_K d}} + \frac{1}{2d}\ln\frac{|\Delta_K|}{d^d} + \ln\frac{\sqrt{d}\mathrm{N}(\mathbf{b}_j)^{1/d}}{\|\mathbf{b}_j\|} + \frac{1}{d}\ln\mathrm{N}(\mathfrak{I}_j)$. By definition of mBKZ reduction, $\mathbf{b}_j/D_j^{1/\beta_K d}$ is a shortest vector in the normalized projected module lattice $\mathcal{M}(\mathfrak{B}_{[\![j;j+\beta_K-1]\!]})^{(1)}$. In other words, we have shown Equation (4).

To conclude, we apply Equation (mGSA) to all $i \in \{1,\ldots,r\}$, giving $\ln\det_{\mathbb{Q}}(\mathcal{M}) = \sum_{i=1}^r\mathbb{E}[\ell_i^K] = r\mathbb{E}[\ell_1^K] - \frac{r(r-1)}{2}\ln\alpha_K$. Hence, $\mathbb{E}[\ell_1^K] = \frac{r-1}{2}\ln\alpha_K + \frac{1}{r}\ln\det_{\mathbb{Q}}(\mathcal{M})$, so $\mathbb{E}[\ell_i^K] = \frac{r+1-2i}{2}\ln\alpha_K + \frac{1}{r}\ln\det_{\mathbb{Q}}(\mathcal{M})$ by Equation (mGSA), as desired. □

### 4.2 Module-Lattice Analog of the Gaussian Heuristic

We approximate the first term $t_1 = \mathbb{E}_{\mathbf{s}}[\ln\|\mathbf{s}\|]$ in Equation (3) by extending the Gaussian Heuristic to module lattices. This is justified if the first projected

module lattice of an $\mathrm{mBKZ}_K^{\beta_K}$-reduced pseudobasis behaves like a random rank-$\beta_K$ module lattice (of the same volume).

Specifically, let $K$ be a number field of degree $d$. For a positive integer $r$, we write $\mathrm{lgh}_K(r)$ for the expected logarithmic first minimum of a random rank-$r$ module lattice over $K$ of unit determinant. In the general case, Minkowski's bound allows us to prove $\mathrm{lgh}_K(r) \leqslant \ln(2) + \mathrm{lgh}_{\mathbb{Q}}(rd)$ for any number field $K$. In the case of a cyclotomic field $K = \mathbb{Q}(\omega_c)$ of conductor $c$ and degree $d = \phi(c)$, a recent study [GSVV24, Theorem 38] proved that the first minimum of a (formally well-defined) random rank-$r$ module lattice of unit determinant is asymptotically concentrated around $(\mu_K/\operatorname{vol}(\mathcal{B}_{rd}))^{1/rd}$, where we recall that $\mu_K$ denotes the number of roots of unity in $K$, and $\mathcal{B}_{rd}$ the $rd$-dimensional Euclidean unit ball.[14]

While this does not predict the exact average, we can attempt to make such a prediction by heuristically 'merging' [GSVV24, Theorem 38] and Heuristic 2. Namely, we use the latter as a baseline for $\mathbb{Q}$, and the former to estimate the gap between $\mathbb{Q}$ and $K$. Because $\mu_{\mathbb{Q}} = 2$, one reaches the following heuristic.

**Heuristic 4 ($\ln \lambda_1$ under the Module-Lattice Gaussian Heuristic [.py]).** *Let $\mathcal{M}$ be a random rank-$r$ module lattice of unit determinant over a number field $K$ of degree $d$. Its expected logarithmic first minimum under the Gaussian Heuristic is given by*

$$\mathbb{E}\left[\ln \lambda_1(\mathcal{M})\right] = \mathrm{lgh}_{\mathbb{Q}}(rd) + \frac{1}{rd} \ln \frac{\mu_K}{2}. \qquad \text{(mGH)}$$

*Hence,* $\mathrm{lgh}_K(r) = \mathrm{lgh}_{\mathbb{Q}}(rd) + \frac{1}{rd} \ln \frac{\mu_K}{2}$.

*Remark 8.* Similar to Heuristic 2, the module-lattice Gaussian Heuristic extends to random module lattices of arbitrary fixed determinant by appropriate scaling.

This results in the following heuristic estimate:

$$t_1 = \mathrm{lgh}_K(\beta_K) = \mathrm{lgh}_{\mathbb{Q}}(\beta_K d) + \frac{1}{\beta_K d} \ln \frac{\mu_K}{2} = \frac{1}{\beta_K d} \left( \ln \frac{\mu_K}{\operatorname{vol}(\mathcal{B}_{\beta_K d})} - \gamma \right)$$

where $\gamma$ denotes the Euler-Mascheroni constant.

We verify this prediction experimentally in Figure 5, and note it to be rather accurate even for relatively small module rank $r$, and increasingly accurate as $r$ grows. We note that even for $K = \mathbb{Q}$ the average is slightly underestimated, but the theorem on which we base the module-lattice Gaussian Heuristic only provides a concentration bound, not an expectation. For $K = \mathbb{Q}$, there also exists a refined estimate for the average [Che13, BSW18], and it would be interesting to extend it to other number fields.

---

[14] We also considered the simpler, specialized Theorem 3 of [GSVV24], but found that it yields poor predictions for even conductors $c$. Recall $\mu_K = 2c$ for odd $c$ and $\mu_K = c$ otherwise.

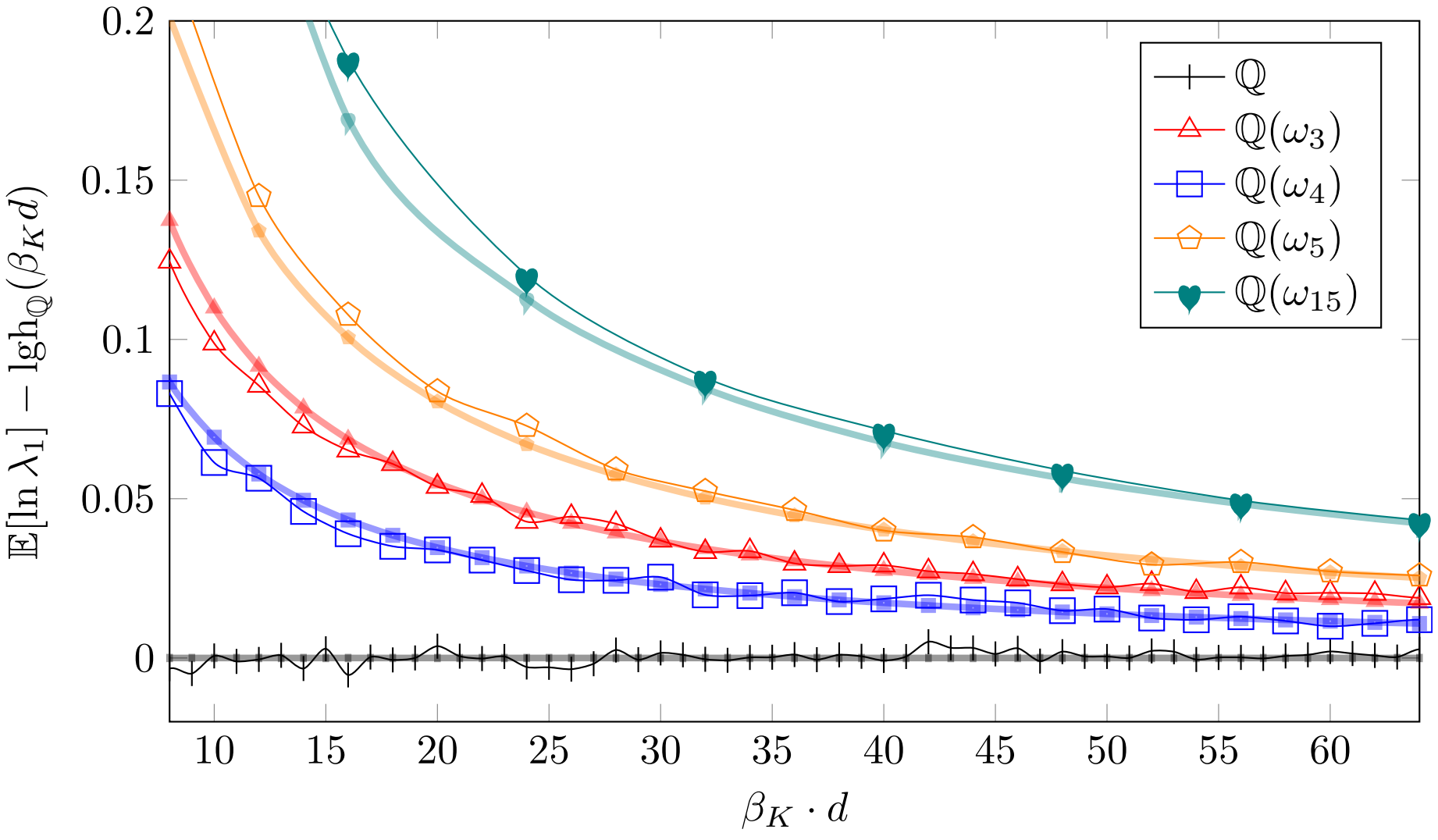

Fig. 5: Logarithmic gap to the Gaussian Heuristic over $\mathbb{Q}$ ($\mathrm{lgh}_{\mathbb{Q}}(\beta_K d)$). Thick lines with small marks are predictions ($\mathrm{lgh}_K(\beta_K)$), thin lines with large marks are experimental average, taken over 1000 samples [.py].

### 4.3 Discriminant Gap

The next term in the formula for $\mathrm{slope}_{\mathbb{Q}}(\mathrm{mBKZ}_K^{\beta})$ is the discriminant gap $t_2 = \frac{1}{2d} \ln\left(\frac{|\Delta_K|}{d^d}\right)$. Ignoring other terms, a smaller discriminant brings the predicted slope closer to 0, thereby contributing to a flatter module-BKZ profile. In the case that $K$ is a cyclotomic field, we have the following explicit formula for its discriminant $\Delta_K$, allowing us to compute the discriminant gap. For $c \in \mathbb{N}$, we write $\mathcal{P}_c$ for the set of distinct prime factors dividing $c$.

**Theorem 9 (Discriminant of Cyclotomic Fields [Was82, Prop. 2.7] [.py]).** *For $c \in \mathbb{N}$, let $\omega_c$ be a primitive $c$-th root of unity. The discriminant $\Delta_K$ of $K = \mathbb{Q}(\omega_c)$ equals $\Delta_K = (-1)^{\frac{\phi(c)}{2}} c^{\phi(c)} \prod_{p \in \mathcal{P}_c} p^{-\frac{\phi(c)}{p-1}}$.*

We obtain the following formula for the discriminant gap $t_2$ of cyclotomic fields, revealing that $t_2$ merely depends on the set $\mathcal{P}_c$ for the field's conductor $c$.

**Lemma 10 (Discriminant Gap Formula).** *For $c \in \mathbb{N}$, let $\omega_c$ be a primitive $c$-th root of unity. Then the discriminant gap $t_2$ of $K = \mathbb{Q}(\omega_c)$ is independent of the exponents in the prime decomposition of $c$, and equals:*

$$t_2 = \frac{1}{2} \sum_{p \in \mathcal{P}_c} \left( \frac{p-2}{p-1} \ln(p) - \ln(p-1) \right).$$

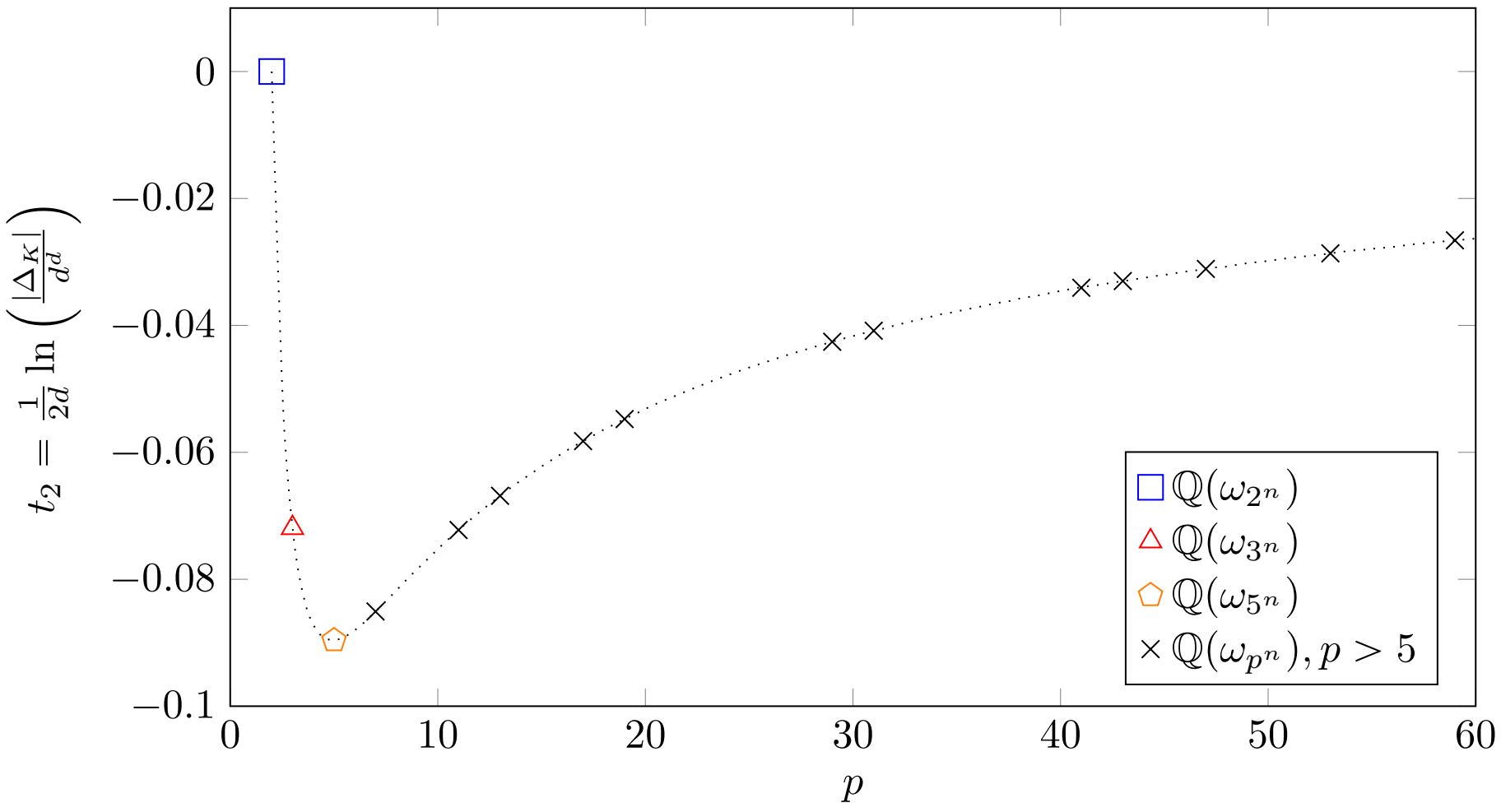

Fig. 6: Discriminant gap $t_2$ for cyclotomic fields of conductor $c = p^n$, where $p$ is a prime and $n \in \mathbb{N}$ [.py].

*Proof.* For $c \in \mathbb{N}$, let $\omega_c$ be a primitive $c$-th root of unity, and define $K := \mathbb{Q}(\omega_c)$ and $d := \phi(c)$. By Theorem 9, $\ln(|\Delta_K|) = d(\ln(c) - \sum_{p\in\mathcal{P}_c} \frac{1}{p-1}\ln(p))$, so by definition the discriminant gap $t_2$ of $K$ equals $t_2 = \frac{1}{2d}\ln(|\Delta_K|) - \frac{1}{2}\ln(d) = \frac{1}{2}\Big(\ln(c) - \ln(d) - \sum_{p\in\mathcal{P}_c}\frac{1}{p-1}\ln(p)\Big)$. By definition of $\phi$, $d = \phi(c) = c\prod_{p\in\mathcal{P}_c}(1-\frac{1}{p})$, so $\ln(c) - \ln(d) = \sum_{p\in\mathcal{P}_c}(\ln(p) - \ln(p-1))$. The result immediately follows. □

In other words, for cyclotomic fields of conductor $c$, we have $t_2 = 0$ if $c$ is a power of 2, and $t_2 < 0$ if $c$ has an odd prime factor. In particular, for a composite conductor $c$, each distinct odd prime factor contributes to decreasing the discriminant gap $t_2$, thereby flattening the predicted module-BKZ profile. However, we remark that the quantity $\frac{p-2}{p-1}\ln(p) - \ln(p-1)$ is minimal for $p = 5$ and is increasing for $p \geqslant 5$. This is illustrated in Figure 6, where we provide the respective values of $t_2$ for prime conductors $c = 5, 7, 11, 3, 13, 17, 19$, listed in increasing order with respect to $t_2$. Asymptotically, for $p \to +\infty$, the contribution of a prime factor $p$ is $\frac{1-\ln(p)}{2p} + O(\frac{1}{p^2})$.

We also numerically checked in `sage` whether subfields of cyclotomic fields could give good values of $t_2$ up to conductor 105 [.py]. They turned out to always be equal or larger than for the original cyclotomic.

### 4.4 Skewness Gap

Next, we consider the skewness gap $t_3 = \mathbb{E}\big[\ln(\frac{\sqrt{d}\mathrm{N}(\mathbf{s})^{1/d}}{\|\mathbf{s}\|})\big]$, where $\mathbf{s}$ is a shortest vector in one of the random projected lattices encountered by module-BKZ. We recall that $t_3$ measures how *skewed* the values $|\sigma(\mathbf{s})|^2$ are for the embeddings

$\sigma$. By construction, this quantity is invariant by scaling the value $\|\mathbf{s}\|$, so we only need to model its direction. Quite naturally, we are tempted to model the direction as being uniform over the $(\beta_K - 1)$-dimensional unit sphere $\mathcal{S}(K_{\mathbb{R}}^{\beta_K})$, forgetting about the fact that $\mathbf{s}$ should be a shortest vector of a module (which certainly constraints its direction [CDPR16]). Conveniently, we model $\mathbf{s} \in K_{\mathbb{R}}^{\beta_K}$ itself as a spherical Gaussian, and denote this modeled skewness gap by $\widetilde{t_3}$. This makes our analysis similar to that of [DvW21, Lemma 4.4], generalized to arbitrary number fields and module ranks.

Let us first express the independent real Gaussian variables more explicitly using the definition of the Euclidean norm. For $\mathbf{s} = (s_1, \ldots, s_{\beta_K})$, we have $\|\mathbf{s}\|^2 = \mathrm{Tr}\left(\langle \mathbf{s}, \mathbf{s} \rangle_K\right) = \sum_{\sigma \in \mathcal{E}} \sum_{j=1}^{\beta_K} \sigma(s_j)\sigma(\overline{s_j})$. For each $j \in [\![1; \beta_K]\!]$, $\sigma_{\mathbb{R}} \in \mathcal{E}_{\mathbb{R}}$, and $\sigma_{\mathbb{C}} \in \mathcal{E}_{\mathbb{C}}^+$, we define $A_{\sigma_{\mathbb{R}},j} := \sigma_{\mathbb{R}}(s_j)$, $B_{\sigma_{\mathbb{C}},j} := \sqrt{2} \cdot \Re(\sigma_{\mathbb{C}}(s_j))$, and $C_{\sigma_{\mathbb{C}},j} := \sqrt{2} \cdot \Im(\sigma_{\mathbb{C}}(s_j))$, giving:

$$\|\mathbf{s}\|^2 = \sum_{j=1}^{\beta_K} \Big( \sum_{\sigma \in \mathcal{E}_{\mathbb{R}}} A_{\sigma,j}^2 + \sum_{\sigma \in \mathcal{E}_{\mathbb{C}}^+} (B_{\sigma,j}^2 + C_{\sigma,j}^2) \Big). \tag{5}$$

On the other hand, we have:

$$\mathrm{N}(\mathbf{s})^2 = \prod_{\sigma \in \mathcal{E}} \sum_{j=1}^{\beta_K} \sigma(s_j)\sigma(\overline{s_j}) = \left( \prod_{\sigma \in \mathcal{E}_{\mathbb{R}}} \sum_{j=1}^{\beta_K} A_{\sigma,j}^2 \right) \left( \prod_{\sigma \in \mathcal{E}_{\mathbb{C}}^+} \frac{1}{2} \sum_{j=1}^{\beta_K} (B_{\sigma,j}^2 + C_{\sigma,j}^2) \right)^2.$$

It is therefore $A_{\sigma,j}, B_{\sigma,j}$, and $C_{\sigma,j}$ in Equation (5) that we model as independent Gaussian variables, say of variance 1. The terms $\sum_{j=1}^{\beta_K} A_{\sigma,j}^2$, $\sum_{j=1}^{\beta_K} B_{\sigma,j}^2$, and $\sum_{j=1}^{\beta_K} C_{\sigma,j}^2$ in the above formulas then follow a $\chi^2$ distribution, and so does $\|\mathbf{s}\|^2$. The expected logarithm of a $\chi^2$ distribution relates to the digamma function, denoted $\psi$, via $\mathbb{E}[\ln \chi_a^2] = \psi(a/2) + \ln 2$, where $a$ is the number of degrees of freedom. We proceed with the calculation [.py], where we use $d = d_{\mathbb{R}} + 2d_{\mathbb{C}}$ to cancel some $\ln 2$ terms and write $\mathcal{U}$ for the uniform distribution over $\mathcal{S}(K_{\mathbb{R}}^{\beta_K})$:

$$\begin{aligned}
\widetilde{t_3} &:= \mathbb{E}_{\mathbf{s} \sim \mathcal{U}} \Big[ \ln \Big( \frac{\sqrt{d}\mathrm{N}(\mathbf{s})^{\frac{1}{d}}}{\|\mathbf{s}\|} \Big) \Big] \\
&= \frac{\ln d}{2} + \frac{d_{\mathbb{R}}\mathbb{E}[\ln \chi_{\beta_K}^2] + 2d_{\mathbb{C}}(\mathbb{E}[\ln \chi_{2\beta_K}^2] - \ln 2)}{2d} - \frac{\mathbb{E}[\ln \chi_{\beta_K d}^2]}{2} \\
&= \frac{\ln d}{2} + \frac{d_{\mathbb{R}}\psi(\beta_K/2) + 2d_{\mathbb{C}}(\psi(\beta_K) - \ln 2)}{2d} - \frac{\psi(\beta_K d/2)}{2}.
\end{aligned} \tag{6}$$

When $K$ is a totally imaginary field, such as a cyclotomic field of conductor $c > 2$, we have $d_{\mathbb{R}} = 0$ and $2d_{\mathbb{C}} = d$. In fact, Equation (6) equals 0 when $K$ is an imaginary *quadratic* field ($d = 2$). Moreover, for $K = \mathbb{Q}$, our model adequately predicts no skewness as well.

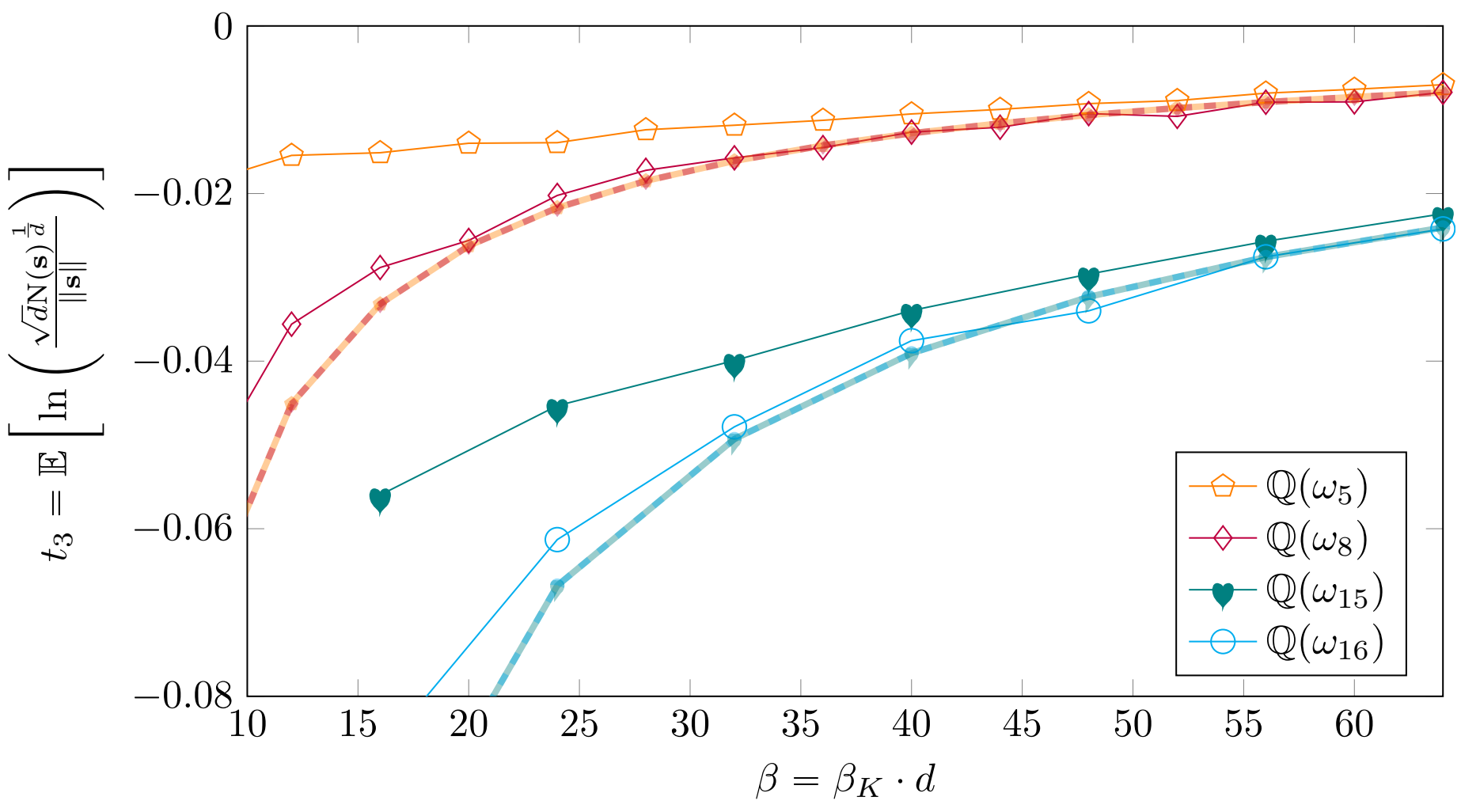

Fig. 7: Skewness gap (term $t_3$). Thick translucent lines with small marks correspond to the prediction $\widetilde{t_3}$ of our model in Equation (6), thin lines with large marks are the experimental average, taken over 1000 samples [.py]. We remark that as the prediction depends only on $d_\mathbb{R}$ and $d_\mathbb{C}$, predictions for different rings can overlap.

Figure 7 suggests that the model and experiments converge for large values of $\beta_K$, but $\widetilde{t_3}$ significantly underestimates the skewness term $t_3$ for smaller $\beta_K$. This means that our model is not exactly accurate. In particular, it does not account for the fact that $\mathbf{s}$ must be a shortest vector of a module lattice, and therefore of $\mathbf{s}\mathcal{O}_K$.

Let us fix $\|\mathbf{s}\| = 1$ by rescaling. In our model, $\mathrm{N}(\mathbf{s})^{1/d}$, has a small probability of being arbitrarily small, for example if the $A_{i,\sigma}$ are small enough for a fixed $\sigma$ and all $i$ (think of points on a sphere of dimension $\beta_K d$ close to a hyperplane of dimension $\beta_K$). However, such a situation is forbidden by Minkowski's bound: if $\mathrm{N}(\mathbf{s})^{1/d}$ gets arbitrarily small, $\det_\mathbb{Q}(\mathbf{s}\mathcal{O}_K)$ does as well, putting an arbitrary small upper bound on the first minimum of the lattice.

### 4.5 Index Gap

Finally, we consider the index gap $t_4 = \frac{1}{d}\mathbb{E}[\ln(\mathrm{N}(\mathfrak{I}))]$. By construction, the pseudobasis is unital, so $\mathcal{O}_K \subseteq \mathfrak{I}$ and thus $\ln(\mathrm{N}(\mathfrak{I})) \leqslant 0$. Let us first clarify that $\ln(\mathrm{N}(\mathfrak{I})) < 0$ does not require $\mathcal{O}_K$ to have non-principal ideals: the shortest vector of a principal ideal may not be a generator. In fact, there are cases where the shortest generator is subexponentially larger than the shortest vector [CDPR16, Lemma 6.2].

We provide a lower bound $\widetilde{t_4}$ on $t_4$ using a density-based argument based on the Dedekind zeta function $\zeta_K$, in a similar way as [ABD16, Sec 2.2], [DPPvW22,

App. A], and [DvW21, Claim 4.2]. We assume here that the random rank-$\beta_K$ modules admit a $K$-basis $\mathbf{B}$ rather than a pseudobasis, and let $\mathbf{v} = \mathbf{B}\mathbf{x}$ be a shortest vector. The ideal $\mathfrak{I}$ constructed in the module-BKZ algorithm is then equal to $\gcd(\{x_i \mathcal{O}_K\}_i)^{-1}$.

Following standard analysis, and assuming that the $x_i$'s are random large elements of $\mathcal{O}_K$, the probability that this gcd is a multiple of an ideal $\mathfrak{a} \subseteq \mathcal{O}_K$ is $\mathrm{N}(\mathfrak{a})^{-\beta_K}$. Writing $\mathcal{D}$ for the distribution of the $\mathfrak{I}$ under our model, and decomposing $\mathfrak{I}$ over the prime ideals of $\mathcal{O}_K$, we obtain [.py]:

$$
\begin{aligned}
\widetilde{t_4} &:= \frac{1}{d}\mathbb{E}_{\mathfrak{I}\sim\mathcal{D}}\left[\ln \mathrm{N}(\mathfrak{I})\right] \\
&= -\frac{1}{d}\sum_{\mathfrak{p}}\sum_{i\in\mathbb{N}} i\cdot\left(\Pr[\mathfrak{I}^{-1}\subset\mathfrak{p}^i] - \Pr[\mathfrak{I}^{-1}\subset\mathfrak{p}^{i+1}]\right)\cdot \ln \mathrm{N}(\mathfrak{p}) \\
&= -\frac{1}{d}\sum_{\mathfrak{p}}\sum_{i\in\mathbb{N}} i\cdot\left(\mathrm{N}(\mathfrak{p})^{-i\beta_K} - \mathrm{N}(\mathfrak{p})^{-(i+1)\beta_K}\right)\cdot \ln \mathrm{N}(\mathfrak{p}) \\
&= -\frac{1}{d}\sum_{\mathfrak{p}}\sum_{i\in\mathbb{N}} \mathrm{N}(\mathfrak{p})^{-i\beta_K} \ln \mathrm{N}(\mathfrak{p}) \\
&= -\frac{1}{d}\sum_{\mathfrak{p}}\frac{\ln \mathrm{N}(\mathfrak{p})}{\mathrm{N}(\mathfrak{p})^{\beta_K}-1} = \frac{1}{d}\frac{\zeta_K'(\beta_K)}{\zeta_K(\beta_K)}
\end{aligned}
\tag{7}
$$

where $\zeta_K$ denotes the Dedekind zeta function of $K$, and $\zeta_K'/\zeta_K$ is its logarithmic derivative [.py]. Here, we used a telescoping identity $\sum_{i=1}^{\infty} i\cdot(x^i - x^{i+1}) = \sum_{i=1}^{\infty} x^i$.

However, over $\mathbb{Q}$, $\mathbb{Q}(\omega_3)$, and $\mathbb{Q}(\omega_4)$, the algebraic norm is an increasing function of the Euclidean norm, so shortest vectors and generators of rank-1 ideals always coincide: it can never be that $\mathrm{N}(\mathfrak{I}) < 1$ despite the above analysis saying it happens with nonzero probability. The previous analysis fails to capture that $\mathbf{v}$ is short, which lowers the probability of finding divisors in this gcd.

The failure of the modeled index gap $\widetilde{t_4}$ is blatant in Figure 8: for small rank $\beta_K$, the model predicts a significant index gap, while for a cyclotomic field of conductor $c \in \{1, 3, 4, 5, 8, 15\}$ the index gap was always trivial over 1000 samples. On the other hand, we did encounter $\mathrm{N}(\mathfrak{I}) < 1$ for conductor 16, but still much less often than predicted. Overall, it seems better to treat this model as a lower bound on the index gap.

Yet, as we discuss in Section 5.1, the modeled index gap $\widetilde{t_4}$ converges exponentially fast to 0 as a function of $\beta_K$ anyway, so this term is actually rather well controlled for large blocksizes.

### 4.6 Conclusion on the Module-BKZ Slope

We now bring together our analysis of the terms $t_1, t_2, t_3, t_4$ and conclude with prediction bounds for the $\mathbb{Q}$-slope of module-BKZ, as illustrated in Figure 1. We

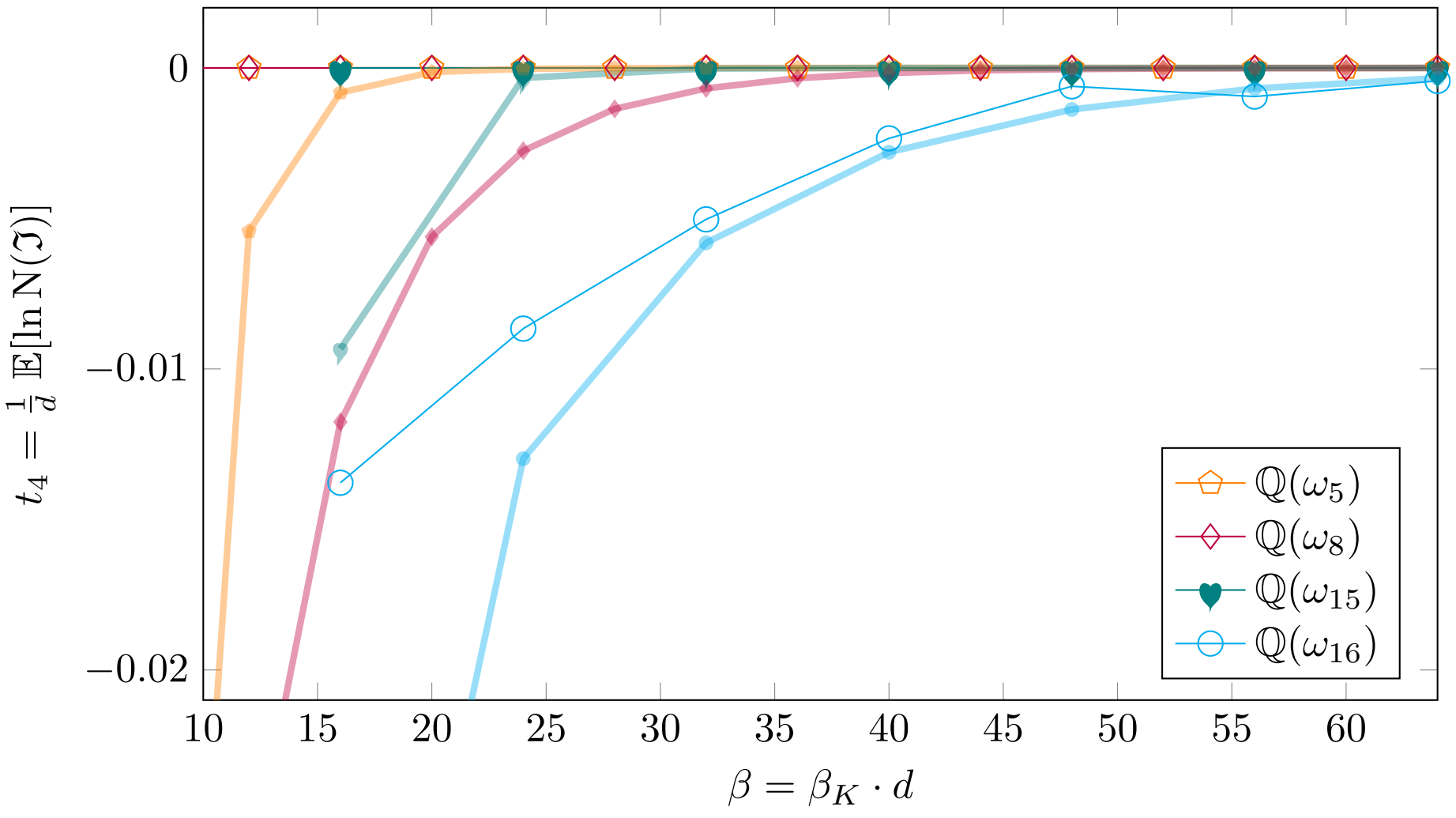

Fig. 8: Index gap (term $t_4$). Thick translucent lines with small marks correspond to the prediction $\widetilde{t_4}$ of our model in Equation (7), thin lines with large marks are the experimental average, taken over 1000 samples [.py].

have shown that for a degree-$d$ number field $K$ the module-BKZ slope satisfies

$$\mathrm{slope}_{\mathbb{Q}}(\mathrm{mBKZ}_K^{\beta_K}) = -\frac{2}{\beta_K d - d}(t_1 + t_2 + t_3 + t_4)$$

for $t_1 = \mathrm{lgh}_{\mathbb{Q}}(\beta_K d) + \frac{1}{\beta_K d} \ln \frac{\mu_K}{2}$, $t_2 = \frac{1}{2d} \ln \frac{|\Delta_K|}{d^d}$, $\frac{\ln d}{2} + \frac{d_{\mathbb{R}} \psi(\beta_K/2) + 2d_{\mathbb{C}}(\psi(\beta_K) - \ln 2)}{2d} - \frac{\psi(\beta_K d/2)}{2} \leqslant t_3 \leqslant 0$, and $\frac{1}{d}\frac{\zeta_K'(\beta_K)}{\zeta_K(\beta_K)} \leqslant t_4 \leqslant 0$. Here, we rely on the module-lattice Gaussian Heuristic for $t_1$, while the lower bounds for $t_3$ and $t_4$ are based on the spherical model and the density-based argument (respectively). In contrast, the formula for $t_2$ does not rely on any heuristic and depends only on $K$.

We implemented these formulas for some cyclotomic fields to predict the module-BKZ $\mathbb{Q}$-slope interval, and compared it against the results from our module-BKZ implementation. This prediction interval and its comparison with practice are shown in Figure 1 in Section 1. Looking at the plots, our prediction interval seems accurate for a rather large range of blocksizes, showing a significant gain for $\mathbb{Q}(\omega_3)$ and $\mathbb{Q}(\omega_{15})$, and a significant loss for $\mathbb{Q}(\omega_8)$, compared to $\mathbb{Q}$. We observe a minor quantitative misfit for $\mathbb{Q}$ and $\mathbb{Q}(\omega_3)$, which might be due to head and tail phenomena that the module-lattice GSA does not account for, but at least qualitatively the gain for $\mathbb{Q}(\omega_3)$ is experimentally confirmed. Such a misfit also appears in unstructured BKZ [YD17,BSW18], and can possibly be overcome by a tail-adapted refinement and simulation (recall Open Questions 1, 2).

## 5 Asymptotic Analysis of the Blocksize Gain

Let $\mathcal{M}$ be a random rank-$r$ module lattice over a degree-$d$ number field $K$. In this section, we compare the predicted BKZ and module-BKZ slopes by determining the blocksize $\beta_{\mathrm{eq}}$ for which $\mathrm{slope}_{\mathbb{Q}}(\mathrm{BKZ}^{\beta}) = \mathrm{slope}_{\mathbb{Q}}(\mathrm{mBKZ}_K^{\beta_{\mathrm{eq}}/d})$, and analyse how it behaves as $\beta \to +\infty$. Concretely, this allows us to quantify the gain or loss of module-BKZ as the difference $\beta_{\mathrm{eq}} - \beta$, as illustrated in Figure 2 in Section 1.

We recall that given an SVP oracle for (unstructured) lattices of dimension $\beta$, the predicted BKZ slope and module-BKZ slope are respectively:

$$\mathrm{slope}_{\mathbb{Q}}(\mathrm{BKZ}^{\beta}) = -\frac{2}{\beta - 1}\,\mathrm{lgh}_{\mathbb{Q}}(\beta),$$
$$\mathrm{slope}_{\mathbb{Q}}(\mathrm{mBKZ}_K^{\beta/d}) = -\frac{2}{\beta - d}(t_1 + t_2 + t_3 + t_4),$$

where the terms $t_1, t_2, t_3, t_4$ are defined with respect to the $K$-rank $\beta_K := \beta/d$.

### 5.1 Asymptotic Behavior of Each Term

First, we detail the asymptotic contribution of each $t_i$ in order to identify the leading asymptotic terms.

**Module-Lattice Gaussian Heuristic $t_1$.** To analyze the first term $t_1$, which is given by $t_1 = \mathrm{lgh}_{\mathbb{Q}}(\beta) + \frac{1}{\beta}\ln\frac{\mu_K}{2}$ under Heuristic 4, we recall that $\mathrm{lgh}_{\mathbb{Q}}(\beta) = \frac{1}{\beta}\big(\ln(2) - \gamma - \ln(\mathrm{vol}(\mathcal{B}_\beta))\big)$ where $\gamma$ denotes the Euler-Mascheroni constant, and $\mathcal{B}_\beta$ the $\beta$-dimensional Euclidean unit ball. Denoting the Gamma function by $\Gamma$, we have $\ln(\mathrm{vol}(\mathcal{B}_\beta)) = \frac{\beta}{2}\ln(\pi) - \ln\Gamma(1+\frac{\beta}{2})$, which implies $\mathrm{lgh}_{\mathbb{Q}}(\beta) = \frac{1}{\beta}\big(\ln(2) - \gamma + \ln\Gamma(1+\frac{\beta}{2})\big) - \frac{1}{2}\ln(\pi)$. Since $\ln\Gamma(z) = z\ln(z) - z - \frac{\ln z}{2} + \frac{\ln(2\pi)}{2} + o(1)$ as $z$ grows, we have $\mathrm{lgh}_{\mathbb{Q}}(\beta) = \frac{\ln(\beta)}{2} - \frac{\ln(2\pi e)}{2} + \frac{\ln(\beta)}{2\beta} + \frac{\ln(4\pi) - 2\gamma}{2\beta} + o\left(\frac{1}{\beta}\right)$, and obtain:

$$t_1 = \frac{\ln(\beta)}{2} - \frac{\ln(2\pi e)}{2} + \frac{\ln\beta}{2\beta} + \frac{\ln(\pi\mu_K^2) - 2\gamma}{2\beta} + o\left(\frac{1}{\beta}\right).$$

**Discriminant Gap $t_2$.** The second term, $t_2 = \frac{1}{2d}\ln\frac{|\Delta_K|}{d^d}$, depends only on $K$, and is therefore constant in $\beta$.

**Skewness Gap $t_3$.** Following our analysis in Section 4.4, we assume that $t_3$ converges to $\widetilde{t_3}$ as $\beta \to +\infty$, where $\widetilde{t_3} = \frac{\ln d}{2} + \frac{d_{\mathbb{R}}\psi(\beta/2d) + 2d_{\mathbb{C}}(\psi(\beta/d) - \ln 2)}{2d} - \frac{\psi(\beta/2)}{2}$ for the digamma function $\psi$. Since the digamma function behaves asymptotically as $\psi(z) = \ln(z) - \frac{1}{2z} + o\big(\frac{1}{z}\big)$, we obtain:

$$t_3 = \frac{1 - d_{\mathbb{R}} - d_{\mathbb{C}}}{2\beta} + o\left(\frac{1}{\beta}\right) = o\left(\frac{\ln\beta}{\beta}\right).$$

**Index Gap $t_4$.** Following our analysis in Section 4.5, we assume that $t_4$ converges to $\widetilde{t_4}$ as $\beta \to +\infty$, where $\widetilde{t_4} = -\frac{1}{d}\sum_{\mathfrak{p}} \frac{\ln \mathrm{N}(\mathfrak{p})}{\mathrm{N}(\mathfrak{p})^{\beta_K}-1}$ with $\mathfrak{p}$ ranging over the prime ideals of $\mathcal{O}_K$ and $\beta_K := \beta/d$. Combining both lemmas in Appendix A then yields $t_4 \geqslant -\sum_p \frac{\ln(p)}{p^{\beta_K}-1}$, where $p$ ranges over the prime integers $p \in \mathbb{N}$ such that $\mathfrak{p} \mid p\mathcal{O}_K$ for some prime ideal $\mathfrak{p} \subset \mathcal{O}_K$. Thus, writing $\mathcal{P}$ for the set of prime integers in $\mathbb{N}$, we obtain $t_4 \geqslant -\sum_{p\in\mathcal{P}} \frac{\ln(p)}{p^{\beta_K}-1} \geqslant -\sum_{p\in\mathcal{P}} \frac{2\ln(p)}{p^{\beta_K}} \geqslant -\sum_{p\in\mathcal{P}} \frac{1}{p^{\beta_K-1}} \geqslant -\frac{1}{2^{\beta_K-1}} - \int_2^\infty \frac{du}{u^{\beta_K-1}}$. In particular, for $\beta_K \geqslant 3$, we have $-\frac{1}{2^{\beta_K-1}} - \frac{1}{(\beta_K-2)2^{\beta_K-2}} \leqslant t_4 \leqslant 0$, so we conclude:

$$t_4 = -\frac{1}{2^{\beta_K-1}} + o\left(\frac{1}{2^{\beta_K}}\right) = o\left(\frac{1}{\beta}\right).$$

### 5.2 Asymptotic Gain Compared to Unstructured BKZ

Finally, we use the previous analysis to show the asymptotic relation between $\beta$ and $\beta_{\mathrm{eq}}$, when $\beta_{\mathrm{eq}}$ is chosen such that $\mathrm{slope}_{\mathbb{Q}}(\mathrm{BKZ}^{\beta}) = \mathrm{slope}_{\mathbb{Q}}(\mathrm{mBKZ}_K^{\beta_{\mathrm{eq}}/d})$.

**Heuristic Claim 3.** *Let $K$ be a number field of degree $d$, and $\beta_{eq} = \beta_{eq}(\beta)$ be such that* $\mathrm{slope}_{\mathbb{Q}}(\mathrm{mBKZ}_K^{\beta_{eq}/d}) = \mathrm{slope}_{\mathbb{Q}}(\mathrm{BKZ}^{\beta})$*. Then, as $\beta \to +\infty$, our heuristic analysis predicts:*

$$\beta_{eq} = \beta + \ln\left(\frac{|\Delta_K|}{d^d}\right)\frac{\beta}{d\ln\beta}\left(1 + \frac{\ln(2\pi e^2)}{\ln\beta} + o\left(\frac{1}{\ln\beta}\right)\right) + d - 1 + o(1).$$

A justification can be found in Appendix B. Note that, in particular, the claim gives:

$$\beta_{\mathrm{eq}} = \beta + \ln\left(\frac{|\Delta_K|}{d^d}\right)\frac{\beta}{d\ln\beta} + o\left(\frac{\beta}{\ln\beta}\right)$$

and when $|\Delta_K| = d^d$:

$$\beta_{\mathrm{eq}} = \beta + d - 1 + o(1).$$

**Acknowledgments.** We thank Kaveh Bashiri and Stephan Ehlen, Nihar Gargava and Vlad Serban, Daniel van Gent, Dustin Moody (and his team), and anonymous reviewers for their valuable feedback. Léo Ducas and Paola de Perthuis were supported by the ERC Starting Grant 947821 (ARTICULATE). Paola de Perthuis was also supported by the NWO Gravitation Project QSC. Lynn Engelberts was supported by the Dutch National Growth Fund (NGF), as part of the Quantum Delta NL program.

# A Decomposition of Prime Integers over Number Fields

The following facts are fairly standard, and are presented in different form in [Neu92, Sam13].

**Lemma 11.** *Let $p \in \mathbb{N}$ be a prime integer, and $K$ a field of degree $d$. Then, $p\mathcal{O}_K$ may be decomposed in a product of at most $d$ prime ideals $\mathfrak{p}$ such that $\mathrm{N}(\mathfrak{p}) = p^k$ for some integer $k$ in $[\![1; d]\!]$.*

*Proof.* Let $p\mathcal{O}_K = \prod_{i\in\mathcal{I}} \mathfrak{p}_i^{\nu_i}$ be the unique decomposition of $p\mathcal{O}_K$ into a product of prime ideals of $\mathcal{O}_K$. Then, $\mathrm{N}(p\mathcal{O}_K) = \mathrm{N}\left(\prod_{i\in\mathcal{I}} \mathfrak{p}_i^{\nu_i}\right) = \prod_{i\in\mathcal{I}} \mathrm{N}(\mathfrak{p}_i)^{\nu_i}$, so $\mathrm{N}(p\mathcal{O}_K) = p^d$ implies $\prod_{i\in\mathcal{I}} \mathrm{N}(\mathfrak{p}_i)^{\nu_i} = p^d$. Hence, for each $i \in \mathcal{I}$, there exists $k_i \in [\![1; d]\!]$ such that $\mathrm{N}(\mathfrak{p}_i) = p^{k_i}$, and $\sum_{i\in\mathcal{I}} k_i\nu_i = d$.

**Lemma 12.** *Let $K$ be a number field of degree $d$. For every prime ideal $\mathfrak{p}$ of $\mathcal{O}_K$, there exists a unique prime integer $p \in \mathbb{N}$ such that $\mathfrak{p} | p\mathcal{O}_K$, and moreover $\mathrm{N}(\mathfrak{p}) = p^k$ for some $k \in [\![1; d]\!]$.*

*Proof.* Let $\mathfrak{p}$ be a prime ideal of $\mathcal{O}_K$. Then, $\mathfrak{p} \cap \mathbb{Z}$ is a prime ideal $\langle p \rangle$ in $\mathbb{Z}$ (*cf.* proof of Theorem 3.1 in [Neu92]). Hence $p\mathcal{O}_K \subset \mathfrak{p}$, and $\mathfrak{p} | p\mathcal{O}_K$. Then, Lemma 11 states that $\mathrm{N}(\mathfrak{p}) = p^k$ for some $k \in [\![1; d]\!]$, and there is no other prime $q \in \mathbb{N}$ with $q \neq p$ such that $\mathfrak{p} | q\mathcal{O}_K$, as then one would have $\mathrm{N}(\mathfrak{p}) = q^\kappa, \kappa \in [\![1; d]\!]$, which would contradict $\mathrm{N}(\mathfrak{p}) = p^k$.

# B Justification of Heuristic Claim 3

In this section, we show how Heuristic Claim 3 is derived from the GSA, the module-lattice Gaussian Heuristic, and our models for the skewness gap and index gap.

*Justification.* By definition of $\beta_{\text{eq}}$, we have:

$$\frac{\operatorname{lgh}_{\mathbb{Q}}(\beta)}{\beta-1} = \frac{t_1+t_2+t_3+t_4}{\beta_{\text{eq}}-d} \tag{8}$$

where the $t_i$ are functions of $\beta_{\text{eq}}$. Our asymptotic analysis from Section 5.1 implies that $t_1 = \operatorname{lgh}_{\mathbb{Q}}(\beta_{\text{eq}}) + o\left(\frac{\ln \beta_{\text{eq}}}{\beta_{\text{eq}}}\right)$, that $t_2 = \frac{1}{2d}\ln\frac{|\Delta_K|}{d^d}$ is constant, and that $t_3$ and $t_4$ are both $o\left(\frac{\ln\beta_{\text{eq}}}{\beta_{\text{eq}}}\right)$. Section 5.1 also yields $2\operatorname{lgh}_{\mathbb{Q}}(\beta) = \ln(\beta) - \ln(2\pi e) + \frac{\ln\beta}{\beta} + o\left(\frac{\ln\beta}{\beta}\right)$. Considering the dominant terms of Equation (8), we obtain $\frac{\ln\beta_{\text{eq}}}{\beta_{\text{eq}}} \sim \frac{\ln\beta}{\beta}$, and in particular $o\left(\frac{\ln\beta_{\text{eq}}}{\beta_{\text{eq}}}\right) = o\left(\frac{\ln\beta}{\beta}\right)$. Therefore:

$$\beta_{\text{eq}} = d + (\beta-1)\cdot A(\beta) \tag{9}$$

where:

$$\begin{aligned} A(\beta) &= \frac{2\operatorname{lgh}_{\mathbb{Q}}(\beta_{\text{eq}}) + 2t_2 + o\left(\frac{\ln\beta}{\beta}\right)}{2\operatorname{lgh}_{\mathbb{Q}}(\beta)} \\ &= 1 + \frac{\ln\left(\frac{\beta_{\text{eq}}}{\beta}\right) + 2t_2 + o\left(\frac{\ln\beta}{\beta}\right)}{\ln(\beta) - \ln(2\pi e) + \frac{\ln\beta}{\beta} + o\left(\frac{\ln\beta}{\beta}\right)}. \end{aligned}$$

As $-\frac{\ln\beta_{\text{eq}}}{\beta_{\text{eq}}} = -\frac{\ln\beta}{\beta} + o\left(\frac{\ln\beta}{\beta}\right)$, applying the $W_{-1}$ branch of the Lambert W function, for $\beta_{\text{eq}} \geqslant 3$, gives: $-\ln\beta_{\text{eq}} = W_{-1}\left(-\frac{\ln\beta}{\beta} + o\left(\frac{\ln\beta}{\beta}\right)\right) = -\ln\beta + o\left(\ln\beta\right)$. Thus, $\ln\frac{\beta_{\text{eq}}}{\beta} = o(\ln\beta)$, and:

$$A(\beta) = 1 + \frac{\ln\left(\frac{\beta_{\text{eq}}}{\beta}\right) + 2t_2}{\delta(\beta)} + o\left(\frac{1}{\beta}\right) = 1 + o(1)$$

for $\delta = \delta(\beta) := \ln(\beta) - \ln(2\pi e) + \frac{\ln\beta}{\beta}$. Replacing $A(\beta)$ in the expression of Equation (9) yields:

$$\frac{\beta_{\text{eq}}}{\beta} = 1 + \frac{\ln\left(\frac{\beta_{\text{eq}}}{\beta}\right) + 2t_2}{\delta(\beta)} + \frac{d-1}{\beta} + o\left(\frac{1}{\beta}\right).$$

Hence:

$$-\delta(\beta)\frac{\beta_{\text{eq}}}{\beta}e^{-\delta(\beta)\frac{\beta_{\text{eq}}}{\beta}} = -\delta(\beta)\ e^{-\delta(\beta)-\varepsilon(\beta)} \tag{10}$$

where $\varepsilon = \varepsilon(\beta) := 2t_2 + (d-1)\ \frac{\ln\beta}{\beta} + o\left(\frac{\ln\beta}{\beta}\right)$. We are looking for $E = E(\beta)$ such that $-\delta\frac{\beta_{\text{eq}}}{\beta} = -\delta - E$. Taking logarithms of Equation (10), and considering $\beta$ large enough for $\delta\frac{\beta_{\text{eq}}}{\beta}$ and $\delta$ to be positive, such an $E$ verifies:

$$-\delta - E + \ln(\delta + E) = -\delta - \varepsilon + \ln\delta$$

which implies:

$$E - \ln\left(1 + \frac{E}{\delta}\right) = \varepsilon. \tag{11}$$

Additionally, as $W_{-1}\left(-\delta\frac{\beta_{\mathrm{eq}}}{\beta}e^{-\delta\frac{\beta_{\mathrm{eq}}}{\beta}}\right) = -\delta\frac{\beta_{\mathrm{eq}}}{\beta} = -\delta - E$, Equation (10) yields: $-\delta - E = W_{-1}\left(-\delta e^{-\delta-\varepsilon}\right)$, and with a first order expansion of $W_{-1}$ in $0^-$: $-\delta - E = -\delta - \varepsilon + \ln\delta + o(\ln\beta)$ which in turn provides $\frac{E}{\delta} = o(1)$. Then, expanding Equation (11) gives:

$$E - \frac{E}{\delta} + o\left(\frac{E}{\ln\beta}\right) = \varepsilon$$

which yields:

$$E = \frac{\varepsilon}{1 - \frac{1}{\ln\beta} + o\left(\frac{1}{\ln\beta}\right)} = \varepsilon\left(1 + \frac{1}{\ln\beta} + o\left(\frac{1}{\ln\beta}\right)\right).$$

Hence:

$$E = 2t_2 + \frac{2t_2}{\ln\beta} + o\left(\frac{1}{\ln\beta}\right).$$

When $t_2 = 0$, we have:

$$E = (d-1)\,\frac{\ln\beta}{\beta} + o\left(\frac{\ln\beta}{\beta}\right).$$

Finally, replacing the expression of $E = 2t_2\left(1 + \frac{1}{\ln\beta} + o\left(\frac{1}{\ln\beta}\right)\right) + (d-1)\,\frac{\ln\beta}{\beta} + o\left(\frac{\ln\beta}{\beta}\right)$ in its definition provides:

$$\frac{\beta_{\mathrm{eq}}}{\beta} = 1 + \frac{2t_2\left(1 + \frac{1}{\ln\beta} + o\left(\frac{1}{\ln\beta}\right)\right) + (d-1)\,\frac{\ln\beta}{\beta} + o\left(\frac{\ln\beta}{\beta}\right)}{\delta(\beta)}$$

and expanding $\frac{1}{1 - \frac{\ln(2\pi e)+o(1)}{\ln\beta}} = 1 + \frac{\ln(2\pi e)}{\ln\beta} + o\left(\frac{1}{\ln\beta}\right)$ gives:

$$\beta_{\mathrm{eq}} = \beta + 2t_2\,\frac{\beta}{\ln\beta}\left(1 + \frac{\ln(2\pi e^2)}{\ln\beta} + o\left(\frac{1}{\ln\beta}\right)\right) + d - 1 + o(1).$$

□

# C BKZ versus mBKZ on module-LWE

While this work is solely focused on the analysis of the slope, its general motivation is cryptanalysis, namely solving module-SIS, module-LWE, module-LIP, and NTRU. A precise quantitative analysis of module-BKZ to solve such problems is left to future work (Open Question 4); yet, we provide experimental evidence that a better slope indeed translates into solving module-LWE more efficiently, at least in some parameter regimes.

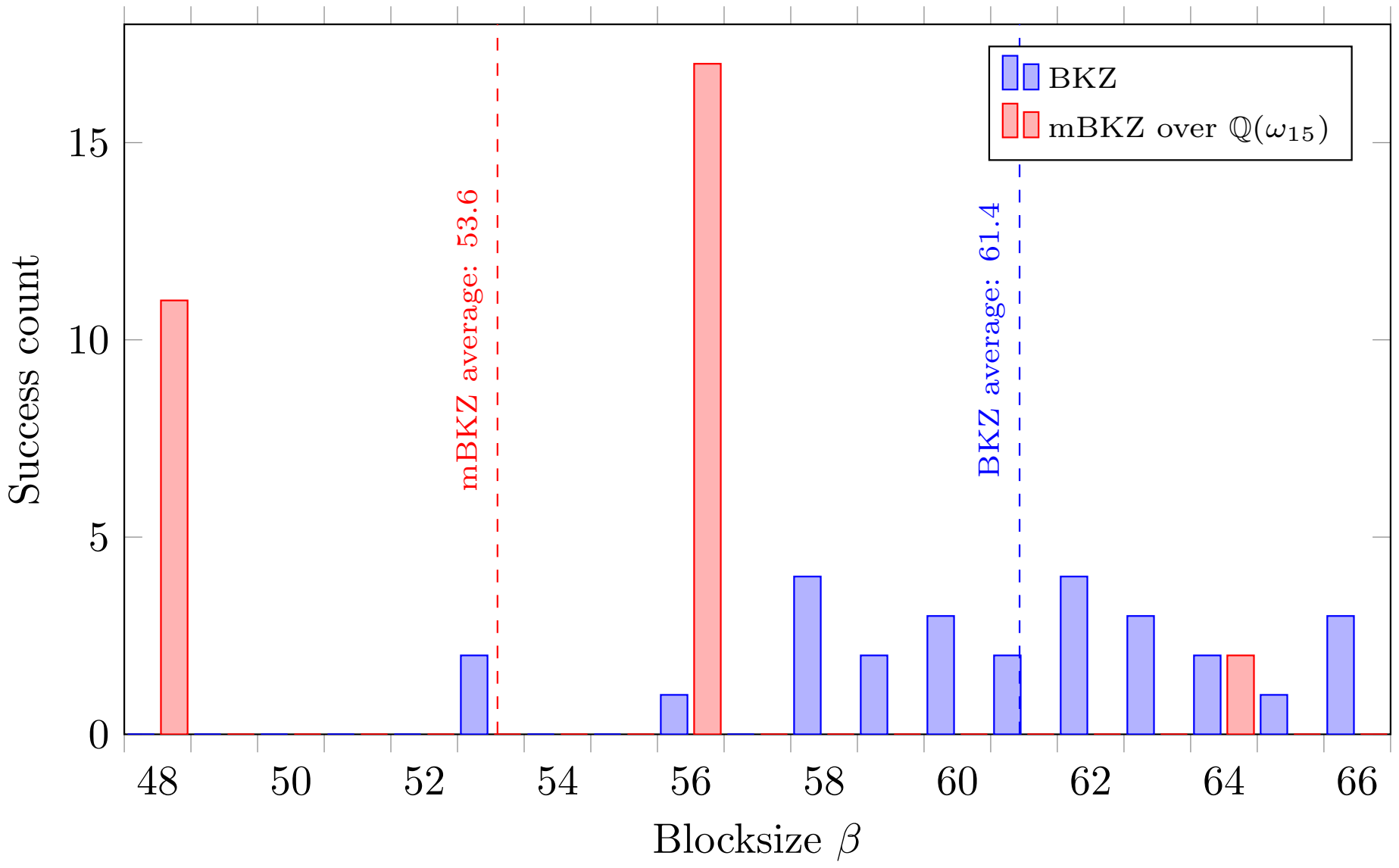

Fig. 9: Counts of successful blocksizes of (m)BKZ to solve a module-LWE problem over $\mathbb{Q}(\omega_{15})$, based on 30 experiments [.py].

More precisely, we constructed a module-LWE instance with small secrets over $\mathbb{Q}(\omega_{15})$ (degree $d = 8$), with secret and error dimension 136, and compared two attacks. The first one (BKZ) [.py] constructs a unique-SVP lattice instance of dimension $273 = 2 \cdot 136 + 1$ using the Kannan embedding, and then runs progressive BKZ until success. The second one [.py] uses a module version of Kannan embedding, leading to a dimension of $280 = 2 \cdot 136 + d$, and runs progressive module-BKZ until success. The results are presented in Figure 9, and show that module-BKZ indeed appears to solve the problem with a significantly smaller average blocksize (53.6 versus 64.1).

Beyond the change of slope, one notes that the module-BKZ attack requires a slightly larger dimension, which may make a small difference. On the other hand, the number of unusually short vectors is now 15 as opposed to 1, which should increase the probability of finding one of them [DDGR20, PV21]. Precise predictions should certainly pay attention to those details.

Furthermore, beyond measuring blocksizes, it should also be noted that each mBKZ tour performs fewer SVP calls than BKZ, by a factor of $d$.